# BATI KARADENİZ HAVZASI'NIN GÜNEY BÖLÜMÜ'NDE AKÇAKOCA-CİDE AÇIKLARINDA SIKIŞMALI TEKTONİK REJİME AİT YENİ BULGULAR

# NEW EVIDENCES OF COMPRESSIONAL TECTONIC REGIME AT THE SOUTHERN PART OF THE WESTERN BLACK SEA BASIN OFFSHORE AKÇAKOCA-CİDE

**Kemal Mert Önal [1*], Emin Demirbağ [2]**

[1] *Sivas Cumhuriyet Üniversitesi, Mühendislik Fakültesi, Jeofizik Mühendisliği Bölümü, 58140, Sivas*

[2] *İstanbul Teknik Üniversitesi, Maden Fakültesi, Jeofizik Mühendisliği Bölümü, 34469, İstanbul*

*Sorumlu yazar / Corresponding author: kmonal@cumhuriyet.edu.tr

---

**Öz**

Batı Karadeniz Havzası'nın güneyinde, 15 Ekim 2016 tarihinde gerçekleşen Karadeniz Depremi ($M_l$=5.0), dikkatleri yeniden havzanın tektonik aktivitesi üzerine çekmiştir. Bu depremin odak mekanizması çözümü, Karadeniz'in Türkiye kıyıları boyunca aletsel olarak kaydedilmiş en büyük depremi olan 03 Eylül 1968 Bartın depreminin ($M_S$=6.6) odak mekanizması çözümüne benzer bir biçimde ters faylanma göstermiş ve bu depremin de bölgede var olan aktif bindirmeye ait diğer bir sismolojik kanıt olduğunu ortaya koymuştur.

Bu çalışmada, Batı Karadeniz Havzası'nın güneyinde, Akçakoca ile Cide arasında kalan bölge açıklarında bulunan şelf ve yamaç bölgeleri altında, sıkışmalı tektonik rejimin etkisi ile oluştuğu düşünülen faylar ve bu fayların aktiviteleri ile oluşan jeolojik yapılar, deniz yansıma sismiği verileri ve kompozit kuyu logu verileri kullanılarak ortaya çıkarılmıştır. Önceki jeolojik çalışmalarda verilen jeolojik kesitler vasıtasıyla kara-kıyı ötesi jeolojik kesitler hazırlanmış ve çalışma alanında karadaki jeolojik özelliklerin deniz altında nasıl devam ettiğine dair bulgular sunulmuştur. Çalışma alanı için elde edilen bu jeolojik kesitler ve sismik göç kesitlerinden ayırt edilen jeolojik yapılar, Batı Karadeniz'de K-G yönlü sıkışmalı tektonik rejimin varlığını destekler niteliktedir.

**Abstract**

The earthquake occurred on October 15, 2016 ($M_l$=5.0) re-attracted attention to the tectonic activity of Western Black Sea Basin. The focal mechanism solution of this earthquake indicates reverse faulting, similar to the Bartın Earthquake of September 3, 1968 ($M_S$=6.6), which is the strongest instrumentally recorded earthquake along the Turkish margin of Black Sea, and reveals another seismological evidence for the active thrusting in the region.

In this study, the fault structures that considered to be formed by the effect of compressional tectonic regime and the structures formed by the activities of these faults beneath the shelf and slope areas between the region offshore Akçakoca-Cide at the southern part of the Western Black Sea Basin revealed by using marine seismic reflection data and the composite well log data. By means of the geological sections in previous geological studies, the land-offshore geological sections were prepared and findings about the continuation of the geological features from land to offshore in the study area were presented. These geological sections developed for the study area and the geological features recognized from the seismic migration sections, support the presence of the N-S directional compressional tectonic regime in Western Black Sea.

---

## GİRİŞ

Karadeniz, 423.000 km$^2$'lik alanı ve 534.000 km$^3$'lük hacmi (Ross, 1974) ile yer kürenin en büyük içsel denizlerinden biri olma özelliğine sahiptir (Şekil 1). Atlantik Okyanusu'na Akdeniz, Ege Denizi ve Türk Boğazlar Sistemi (Marmara Denizi, İstanbul ve Çanakkale Boğazları) aracılığıyla bağlı olan Karadeniz'in ortalama su seviyesi, Ege Denizi'nin su seviyesinden yaklaşık olarak 55 cm daha yüksektedir (Alpar ve Yüce, 1998). Karadeniz, kuzeyindeki Azak Denizi'ne, derinliği 5 m'den az olan Kerç Boğazı ile bağlıdır (Ross ve diğ., 1974). Güneyden Türkiye, batıdan Romanya ve Bulgaristan, kuzey ve doğudan da Ukrayna, Rusya ve Gürcistan'ın çevrelediği oval bir havza olan Karadeniz, Andrusov ve Archangelsky Sırtlarının oluşturduğu Orta Karadeniz Sırtı tarafından Batı ve Doğu Karadeniz Havzalarına ayrılmaktadır. Karadeniz'in, gerek derin yapısının gerekse jeolojik özelliklerinin araştırılması amacıyla, 19. yüzyılın sonlarından başlayarak günümüze kadar uzanan zaman dilimi içerisinde çok sayıda jeolojik, jeofiziksel, jeokimyasal ve oşinografik çalışma gerçekleştirilmiştir.

Jeofiziksel çalışmalardan elde edilen veriler, 1960'lardan itibaren hızla gelişim gösteren tektonik bilimi için giderek artan bir öneme sahip hale gelmiştir. Sismik veriler ve bunun yanı sıra manyetik ve gravite verileri, yerbilimcilerin gözlemlerine kritik üçüncü uzamsal boyutu ekleyerek, derindeki büyük ölçekli yapıların geometrileri hakkında detaylı bilgiler elde edilmesini sağlamaktadır. Sismolojik çalışmalar ve bunların yanı sıra manyetizma ve paleomanyetizma çalışmaları da, küresel tektonik düzenlemelerin yeniden yapılabilmesi için gerekli olan, levhaların geçmiş ve günümüz geometrileri ve hareketleri hakkında veriler sunmaktadır (Moores ve Twiss, 1995).

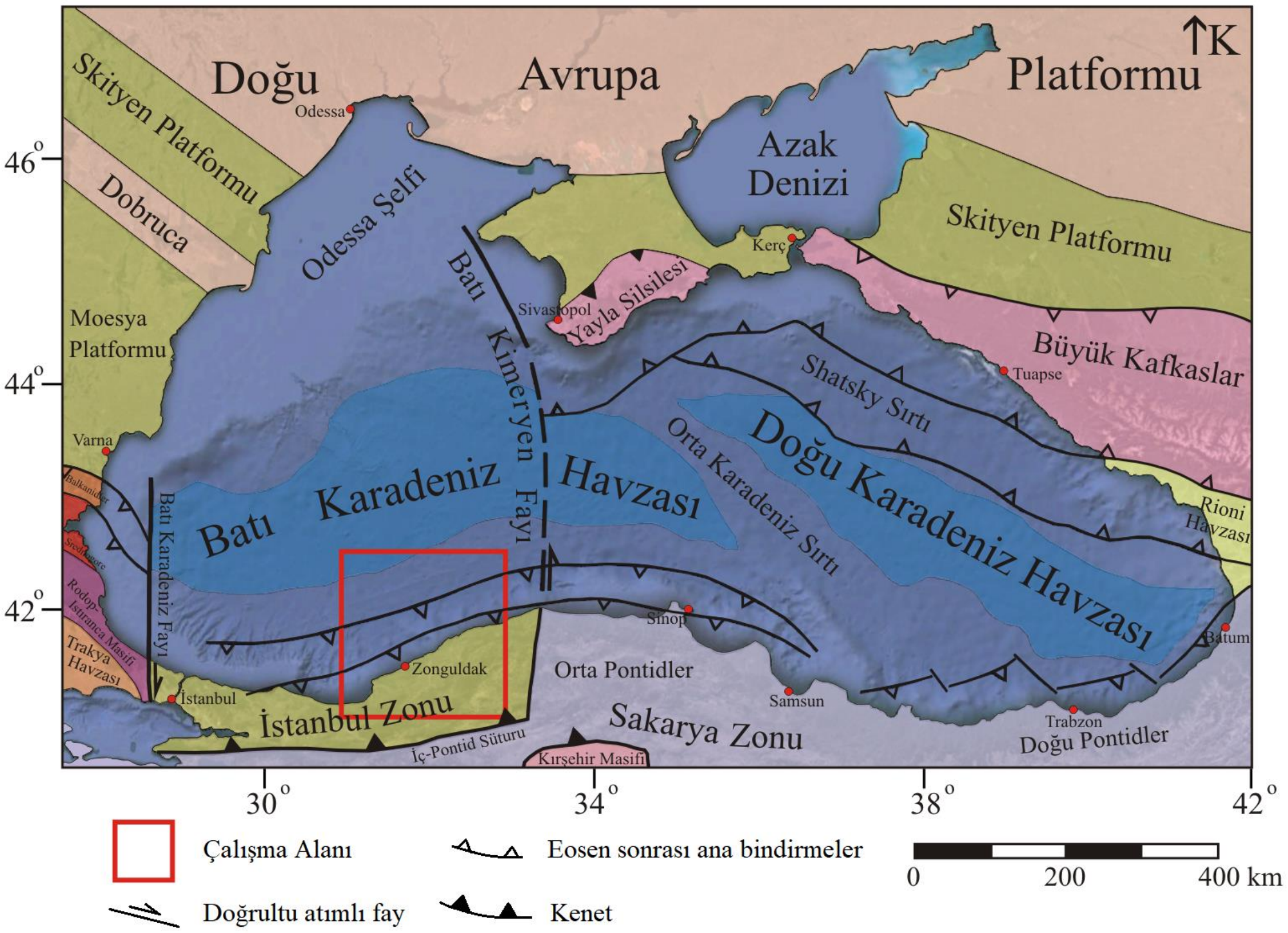

**Şekil 1.** Karadeniz ve çevresinin tektonik haritası (Okay ve diğ. (1994) ve Okay ve Tüysüz (1999)'den yararlanılarak oluşturulmuştur).

**Figure 1.** Tectonic map of Black Sea Basin and the surrounding region (modified after Okay et al. (1994) and Okay and Tüysüz (1999)).

Sismik yansıma yöntemi, denizlerde taban altındaki derin yapıların ve jeolojik özelliklerin incelenmesinde en etkin biçimde kullanılan jeofiziksel yöntemdir. Batı Karadeniz'in Türkiye kıyıları açıklarındaki şelf ve yamaç bölgelerinde petrol arama firmaları tarafından ekonomik amaçlı olarak gerçekleştirilen iki boyutlu (2B) ve üç boyutlu (3B) sismik çalışmalar, akademik çevrelerin kullanımına oldukça kısıtlı bir şekilde açıktır. Bu çalışmaların sayısıyla kıyaslanamayacak kadar az olsa da, bu bölgedeki tektonik çalışmalara ışık tutan bilimsel amaçlı sismik yansıma çalışmalarının sayısında özellikle 1990'lı yılların ortalarından itibaren önemli bir artış gözlenmektedir. TPAO ve diğer petrol arama firmaları tarafından toplanan yoğun miktardaki sismik verilerin oldukça az bir bölümü, bu bilimsel amaçlı çalışmalara katkı sağlayacak şekilde yayınlanarak akademik ortamda paylaşılmıştır (örn. Robinson ve diğ., 1995; Robinson ve diğ., 1996; Yiğitbaş ve diğ., 2004; Menlikli ve diğ., 2009; Tari ve diğ., 2015; Tari ve diğ., 2016; İşcan ve diğ., 2017; Sipahioğlu ve Batı, 2018; Ocakoğlu ve diğ., 2018). Batı Karadeniz'in Türkiye kıyıları açıklarında günümüze kadar, özellikle kütle kayması, gaz hidratlar ve tabana benzeyen yansıtıcı (Bottom Simulating Reflector-BSR), sedimantasyon, deniz-seviyesi değişimleri ve Messiniyen olayları gibi farklı konular üzerine odaklanmış, detaylı sismik kesitler sunan birçok önemli bilimsel çalışma gerçekleştirilmiştir (Konuk ve diğ., 1991; Alpar ve diğ., 1997; Demirbağ ve diğ., 1999; Algan ve diğ., 2002; Aksu ve diğ., 2002; Hiscott ve Aksu, 2002; Gillet ve diğ., 2003; Alpar ve Gainanov, 2004; Kuşçu ve diğ., 2004; Gillet ve diğ., 2007; Flood ve diğ., 2009; Lericolais ve diğ., 2011; Okay ve diğ., 2011; Dondurur ve diğ., 2012; Dondurur ve diğ., 2013; Nikishin ve diğ., 2015a; Nikishin ve diğ., 2015b; Suc ve diğ., 2015).

Tektonik konulu çalışmalarda, depremler ve heyelanlar gibi zarar verici tektonik olayların zaman ve yerlerinin önceden tahmin edilmesi için çeşitli yollar aranmaktadır. Neotektonik açıdan yararlı olabilmesi için bu önceden tahminlerin, can ve mal kaybını azaltmak için yeterli doğrulukta olma zorunluluğu vardır. Bu önceden tahmin tekniklerinin araştırılması, aktif tektonik özelliklerin araştırılmasını da kapsamaktadır (Moores ve Twiss, 1995). Batı Karadeniz Havzası'nın güneyinde, 15 Ekim 2016 tarihinde yerel saatle 11:18'de gerçekleşen Karadeniz depremi ($M_l$=5.0) (KRDAE, 2016), dikkatleri yeniden Karadeniz'in tektonik aktivitesi konusu üzerine çekmiştir (Şekil 2).

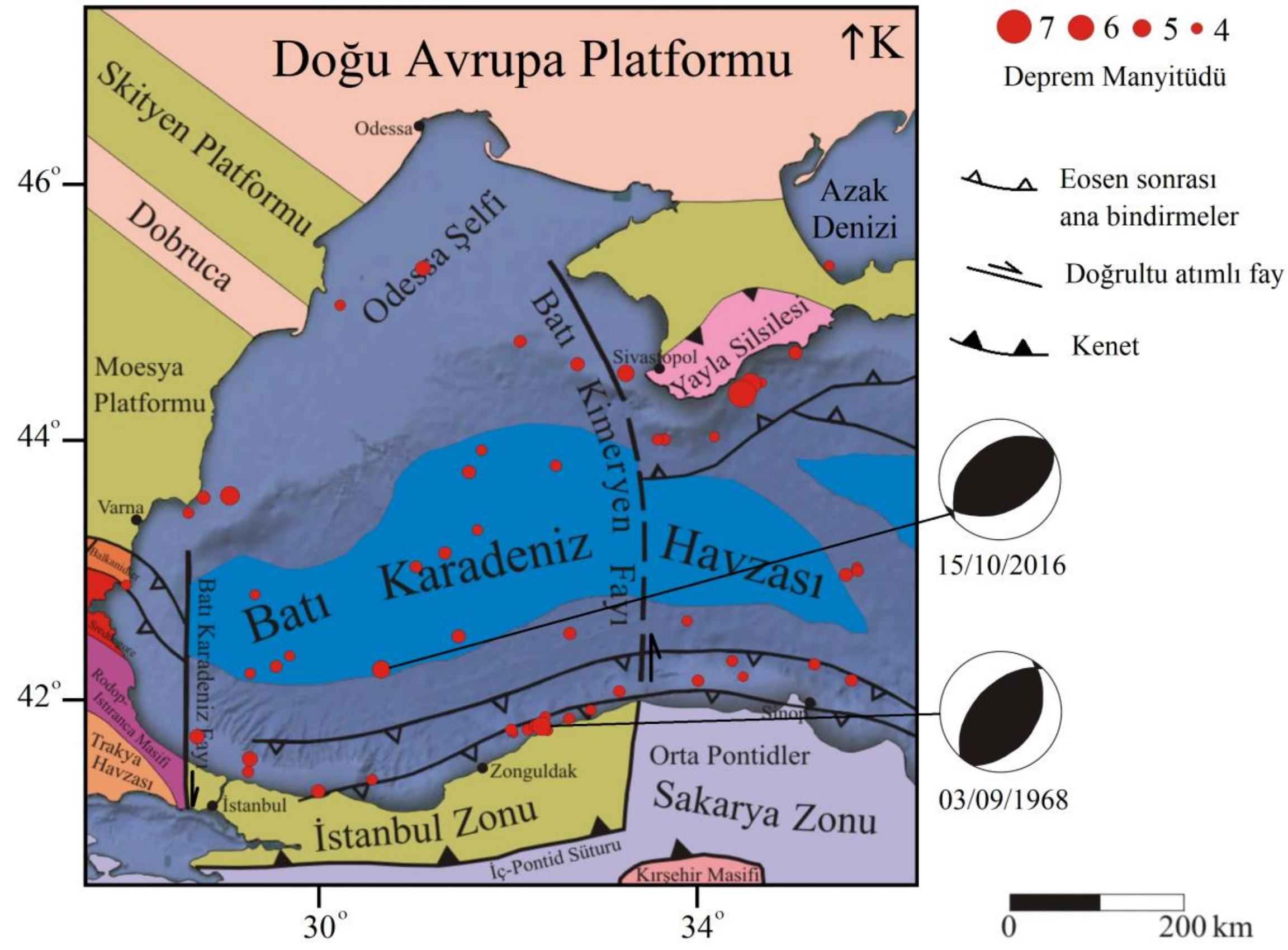

**Şekil 2.** Batı Karadeniz Havzası ve çevresinin tektonik haritası üzerinde 1900-2017 yılları arasında M ≥ 4.0 olarak kıyıdan açıkta gerçekleşen depremler (ISC, 2017). 15.10.2016 (Url-1) ve 03.09.1968 (Taymaz ve diğ., 1999) depremlerinin odak mekanizması çözümlerinin verildiği harita, Okay ve diğ. (1994) ve Okay ve Tüysüz (1999)'den yararlanılarak hazırlanmıştır.

**Figure 2.** Offshore earthquakes (M ≥ 4.0) from 1900 to 2017 (ISC, 2017) on tectonic map of Western Black Sea Basin and the surrounding region. The map including fault plane solutions of 15.10.2016 (Url-1) and 03.09.1968 (Taymaz et al., 1999) earthquakes prepared by utilising Okay et al. (1994) and Okay and Tüysüz (1999).

İstanbul'un 195 km kuzeydoğusunda, Zonguldak'ın ise 124 km kuzeybatısında kalan ve yüzeyden 11.4 km derinde olan bir merkez üssüne sahip olarak meydana gelen Karadeniz depremi, yaklaşık olarak 7-8 saniye sürmüştür. İstanbul, Kocaeli, Düzce, Sakarya, Zonguldak ve hatta Bulgaristan'ın Varna ve Burgaz şehirlerinde hissedilen bu deprem, Karadeniz'in Türkiye kıyılarına yakın bölgede aletsel olarak kaydedilebilen nadir büyük depremlerden biridir. 15 Ekim 2016 Karadeniz Depremi'nin moment tensör çözümünden elde edilen odak mekanizması çözümü (Url-1), Karadeniz'in Türkiye kıyılarında aletsel olarak kaydedilmiş en büyük depremi olan 03 Eylül 1968 Bartın depreminin ($m_b$=5.7, $M_S$=6.6) (Alptekin ve diğ., 1986) odak mekanizması çözümüne benzer bir biçimde ters faylanma göstererek, bu depremin de bölgede var olan aktif bindirmeye ait diğer bir sismolojik kanıt olduğunu ortaya koymuştur.

Karadeniz'e kıyısı olan İstanbul, Kocaeli, Düzce, Sakarya, Zonguldak ve Bartın, nüfus bakımından Türkiye'nin yaklaşık olarak % 23'lük bir bölümünün (toplamda yaklaşık olarak 18 milyon 500 bin insan) yaşadığı şehirlerdir. Ayrıca İstanbul ve Kocaeli, günümüzde Türkiye'nin en önemli iki sanayi şehri olma özelliğini korumaktadır. Karadeniz'in Türkiye kıyıları açıklarında meydana gelebilecek büyük depremler sonucu oluşacak tektonik afetler ve bu afetlerin doğuracağı kötü sonuçlar, yerbilimcileri bölgede, özellikle de kıyı açıklarındaki şelf ve yamaç bölgelerinin altındaki derin yapının ve jeolojik özelliklerin ortaya çıkarılabilmesi için, daha detaylı bilimsel çalışmalar gerçekleştirmeye yönlendirmektedir.

Bu çalışma, Karadeniz kıyısında Akçakoca ile Cide arasında kalan bölge açıklarında bulunan deniz yamacı altında, muhtemel aktif faylar sonucu oluşmuş jeolojik yapıların (tabakalanmalar, kıvrımlar, bindirme ile ilişkili yapılar vb.) özelliklerinin deniz yansıma sismiği verileri kullanılarak belirlenmesi ve karadaki jeolojinin deniz altında nasıl devam ettiğine dair bilgilerin ortaya konulması ile ilgili ilksel sonuçları vermektedir.

## BATI KARADENİZ HAVZASI'NIN TEKTONİK ÖZELLİKLERİ

Türkiye'nin kuzeyinde, Balkanlar ve Kafkaslar arasında uzanarak Karadeniz'in güney kıtasal sınırını oluşturan Pontidler (Ketin, 1966), Alpin dağ kuşağının önemli bir bölümünü oluşturmaktadır (Okay ve Tüysüz, 1999; Tüysüz, 2018). Bulgaristan'ın Srednogore magmatik kuşağını Kafkaslara bağlayan bu magmatik kuşağın, Tetis Okyanusu'nun kuzey kolunun kuzeye doğru yitiminin bir sonucu olduğu üzerine genel bir fikir birliği mevcuttur (Boccaletti ve diğ., 1974; Peccerillo ve Taylor, 1975; Boccaletti ve diğ., 1978; Manetti ve diğ., 1979; Şengör ve Yılmaz, 1981; Yılmaz ve diğ., 1997; Tüysüz, 1999; Tüysüz ve diğ., 2012; Tüysüz ve diğ., 2016; Keskin ve Tüysüz, 2018).

Karadeniz'in kuzeyden sınırladığı Pontidler, batıdan doğuya doğru, Istranca, İstanbul ve Sakarya zonları (Okay, 1989) olmak üzere, üç farklı kıtasal parçadan meydana gelmektedir (Keskin ve Tüysüz, 2018). Pontidler, coğrafi açıdan Doğu, Orta ve Batı Pontidler bölümlerine ayrılmaktadır. Batı Pontidler bu kuşağın İstanbul ile Kastamonu arasındaki, Orta Pontidler Kastamonu ile Samsun arasındaki, Doğu Pontidler ise Samsun'dan daha doğudaki kesimini içermektedir (Tüysüz, 1993). Ketin (1966)'in İzmir-Ankara-Erzincan ofiyolitik kuşağı olarak tanımladığı ve daha sonra Şengör ve Yılmaz (1981) tarafından İzmir-Ankara-Erzincan Kenedi olarak yeniden adlandırılan kuşak, Pontidleri güneyden sınırlandırmaktadır. İstanbul ve Sakarya zonlarını, daha sonradan Kuzey Anadolu Fay Zonu (KAFZ) tarafından değişikliğe uğratılmış veya yok edilmiş olan ve Şengör ve Yılmaz (1981) tarafından İç-Pontid Kenedi olarak adlandırılan kuşak ayırmaktadır (Tüysüz, 2018).

Karadeniz'in Pontid magmatik kuşağının kuzeyine doğru bir yay-ardı ve/veya yay-içi havza olarak açılmış olması üzerine genel bir fikir birliği vardır (Letouzey ve diğ., 1977; Zonenshain ve Le Pichon, 1986; Finetti ve diğ., 1988; Görür, 1988; Manetti ve diğ., 1988; Tüysüz ve diğ., 2012; Tüysüz ve diğ., 2016; Keskin ve Tüysüz, 2018). Ayrıca, Karadeniz Havzası'nı oluşturan Batı ve Doğu Karadeniz Havzaları'nın kıtasal Andrusov ve Archangelsky sırtları tarafından ayrıldıkları ve bundan dolayı da farklı açılma zamanları ve mekanizmalarına sahip oldukları üzerine genel bir uzlaşma mevcuttur (Okay ve diğ. 1994; Robinson ve diğ., 1996; Spadini ve diğ., 1997; Shillington ve diğ., 2008; Shillington ve diğ., 2009; Stephenson ve Schellart, 2010; Munteanu ve diğ., 2011; Okay ve diğ., 2013; Keskin ve Tüysüz, 2018). Batı ve Doğu Karadeniz Havzaları'nın her ikisinde de, Kretase'den günümüze kadar geçen zaman içinde çökelmiş, Batı Karadeniz Havzası'nda 14 km'den kalın ve Doğu Karadeniz Havzası'nda ise 12 km'den daha az kalınlıkta olan, kalın birer sedimenter örtü bulunmaktadır (Zonenshain ve Le Pichon, 1986; Finetti ve diğ., 1988; Okay ve diğ., 1994; Tüysüz, 1999). Her iki havzanın geometrik yapıları açısından bakıldığında, Doğu Karadeniz Havzası aşağı yukarı simetrik olan bir yapıya sahip iken, Batı Karadeniz Havzası belirgin bir asimetriye sahiptir. Batı Karadeniz Havzası'nın güney bölümü dik eğimli ve dar iken kuzey bölüm daha hafif eğimli bir yapıya sahiptir (Nikishin ve diğ., 2015a).

Batı Karadeniz Havzası'nın güney sınırının bir bölümünü oluşturan İstanbul Zonu (Okay, 1986), Kretase'ye kadar, Moesya Platformu ve Kırım arasında bulunan Odessa şelfi boyunca konumlanmaktaydı. Erken Kretase'de İstanbul Zonu riftleşerek, Batı Karadeniz ve Batı Kimeryen Fayları boyunca güneye doğru hareket etmeye başlamış ve Batı Karadeniz Havzası da, bir yay-ardı havza olarak, güneye doğru hareket eden İstanbul Zonu'nun arkasından açılmaya başlamıştır. (Okay ve diğ., 1994; Tüysüz, 2018). İstanbul Zonu, muhtemelen Geç Albiyen sırasında günümüzdeki konumuna yerleşmiştir. Senomaniyen sırasında, İstanbul Zonu ile Sakarya Zonu'nun İç-Pontid Kenedi boyunca çarpışması, İstanbul Zonu'nun yükselmesine sebep olmuştur (Tüysüz, 2018). Çarpışma sonrası sıkışma kuzeye doğru gelişmiş ve Orta Eosen'den itibaren tüm Pontidler boyunca bir kıvrım ve bindirme kuşağının gelişmesine yol açmıştır. (Sunal ve Tüysüz, 2002; Keskin ve Tüysüz, 2018). Bu kıvrım ve bindirme kuşağı, kuzey ve güney verjanslı bindirmeler ve yoğun kıvrımlı Eosen ve Eosenden daha yaşlı çökellerden oluşmaktadır (Keskin ve Tüysüz, 2018).

Batı Pontidler, Geç Eosen ve Erken Oligosen zamanlarında kıvrımlanmış ve Karadeniz'e doğru bindirmişlerdir (Okay ve diğ., 2001; Sunal ve Tüysüz, 2002; Cavazza ve diğ., 2011; Nikishin ve diğ., 2015a). Bu kıvrım-ve-bindirme kuşağı tabanda bir sıyrılma yüzeyi ile ayrılmaktadır (Sunal ve Tüysüz, 2002; Nikishin ve diğ., 2015a). Bölgede yer alan bindirme ve sıyrılma fayları derinden başlayıp yüzeye, daha genç çökellerin olduğu sığlığa doğru gelişmişlerdir. Sarplık (ramps) ve düzlüklerden (flats) oluşan bindirmeler, bölgede yer alan litolojik özelliklere göre gelişmişlerdir (Sunal, 1998; Sunal ve Tüysüz, 2001; Sunal ve Tüysüz, 2002). Örneğin kireçtaşı gibi dayanımlı birimlerde sarplıklar oluşurken daha düşük dayanımlı birimlerde, yani killi birimlerde genellikle düzlükler, başka bir ifadeyle sıyrılma fayları oluşmaktadır. Bu oluşum, tipik kıvrım ve bindirme kuşaklarında (fold-and-thrust belts) gelişen ve birimlerin dayanım farklılıklarından (competence difference) kaynaklanan bir yapıdır (Mitra, 1986; Mitra 1990). Kıvrım ve bindirme yapıları ile sıyrılma yüzeyi Maykop çökelleri tarafından örtülmüştür (Nikishin ve diğ., 2015a). Batı Pontidlerin Apatit fizyon izi termokronolojisi, Pontidlerdeki ana yükselimin Geç Lütesiyen-Erken Rupeliyen süresince gerçekleştiğini göstermektedir (Cavazza ve diğ., 2011; Nikishin ve diğ., 2015a).

Sıkışma, bindirme ve yükselme olaylarının varlığı sonucu Batı Karadeniz Havzası'nın güney ve güneybatı bölümleri boyunca oluşmuş Paleosen ve Eosen türbidit havzaları, 2B sismik veriler kullanılarak oluşturulan kesitler üzerinden ayırt edilerek haritalandırılmıştır (Nikishin ve diğ., 2015a; Nikishin ve diğ., 2015b). Oligosen-Miyosen zamanları, Maykop çökelleri için sediman kaynağı olan bir yükselme ve aşınma (erozyon) periyodu olarak gelişmiştir (Keskin ve Tüysüz, 2018).

Batı Karadeniz Havzası'nın güney kıyıları boyunca yükselme olaylarının varlığının, kıyı boyunca yükselmiş olan denizel taraçalara vurgu yapılarak desteklendiği birçok çalışma mevcuttur (örn. Erinç ve İnandık, 1955; İnandık, 1956; Wedding, 1968; Ketin ve Abdüsselamoğlu, 1969; Barka ve Sütçü; 1993; Gökaşan, 1996; Koral 2007a; Koral 2007b; Yıldırım ve diğ., 2011; Yıldırım ve diğ., 2013; Ilgar, 2014; Yıldırım ve diğ., 2019). Ayrıca Batı Karadeniz Havzası'nın güney kıyıları boyunca sıkışmalı güncel bir tektonik aktivitenin varlığını destekleyen birçok çalışma bulunmaktadır (örn. Alptekin ve diğ., 1986; Gökaşan, 1996; Sunal ve Tüysüz, 2002; İşcan ve diğ., 2017).

Tüysüz (1999) tarafından verilen, Batı Pontidlerin başlıca tektonik unsurlarını içeren haritanın, bu çalışma kapsamında incelenen bölgeyle sınırlandırılmış hali Şekil 3'te verilmektedir. İstanbul Zonu'nun Paleozoyik temeli İstanbul-Akçakoca kuşağında, Sünnice masifinde ve Zonguldak havzasının kuzeyinde görülmektedir (Tüysüz, 1999). Devrek havzasının Üst Kretase-Eosen birimleri tarafından örtülen Zonguldak havzası, Geç Barremiyen ile Senomaniyen arasında çökelmiş karbonat ve kırıntılıları içermektedir. Ulus havzası, Zonguldak havzası gibi Kretase çökelleri ile doldurulmuş olmasına rağmen bazı farklılıklar sunar. Zonguldak havzasında kıyı fasiyeslerin yaygın olmasına karşılık Ulus havzasında yaşıt birimler daha çok türbiditik nitelikli havza içi çökellerden oluşmaktadır (Tüysüz ve diğ., 2004). Ulus havzası güney doğusundaki Safranbolu havzasının Tersiyer çökelleri üzerine bindirmiştir. Cide yükselimi, Zonguldak ve Ulus havzası ile benzer kayaçlardan oluşmaktadır. İstanbul zonu güneyde İç-Pontid Kenedi ile sınırlandırılmaktadır (Tüysüz, 1999).

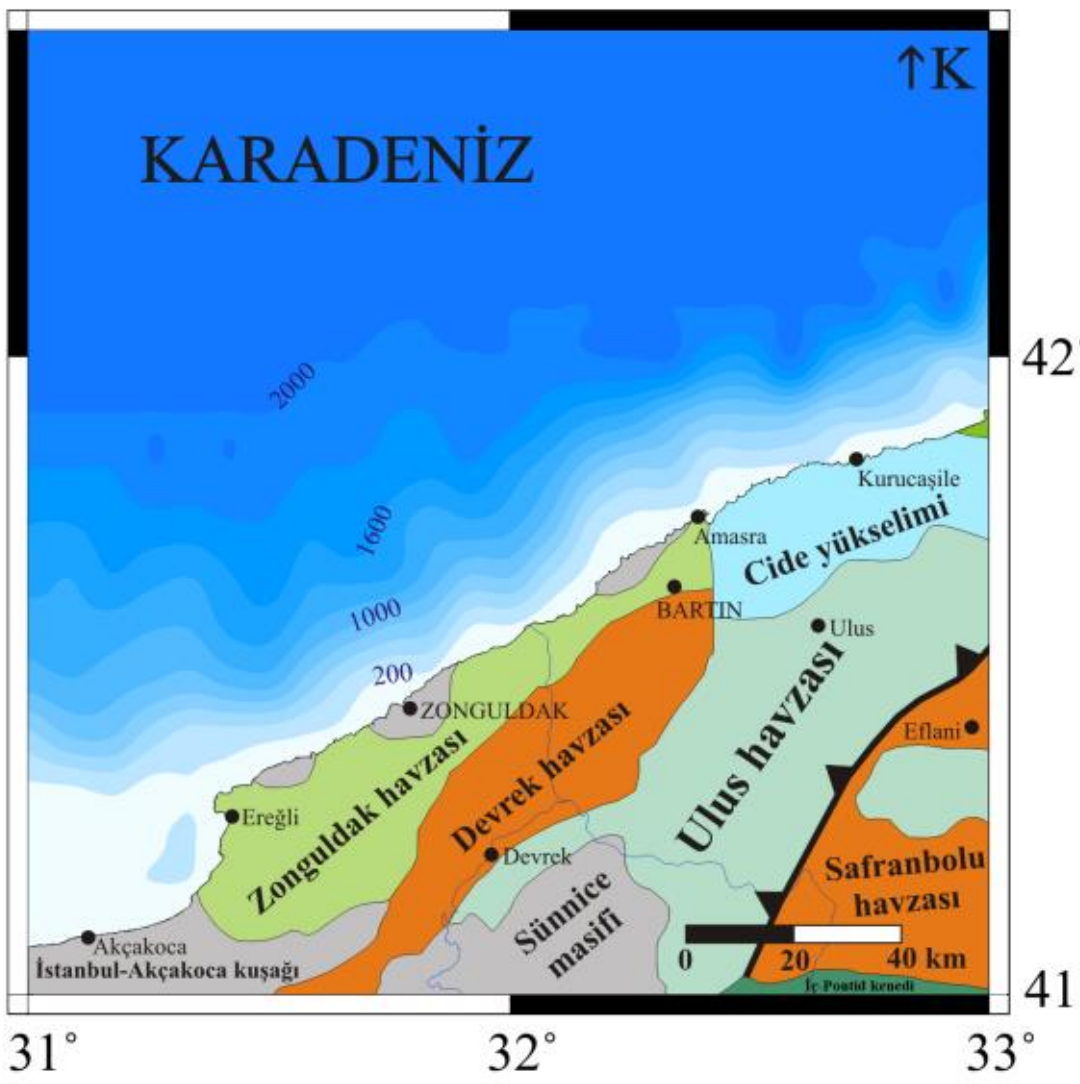

**Şekil 3.** Çalışma alanı başlıca tektonik unsurları (Tüysüz (1999)'den yararlanılarak hazırlanmıştır. Haritanın batimetrisi, GEBCO (IOC, IHO ve BODC, 2003) verileri ile GMT (Wessel ve diğ., 2013) kullanılarak hazırlanmıştır).

**Figure 3.** Main tectonic features of the study area (prepared by utilising Tüysüz (1999). Bathymetry of the map prepared by using GMT (Wessel et al., 2013) by the data from GEBCO (IOC, IHO and BODC, 2003)).

## BATI KARADENİZ HAVZASI'NIN DEPREMSELLİĞİ

Karadeniz'in Türkiye kıyılarında aletsel olarak kaydedilmiş en büyük deprem, 03 Eylül 1968'de meydan gelen Bartın depremidir ($m_b$=5.7, $M_S$=6.6). Bartın depremi, sağ yanal doğrultu atımlı Kuzey Anadolu Fayı (KAF) boyunca meydana gelen depremlerden farklılık göstermiştir (Alptekin ve diğ., 1986). Ters faylanma gösteren odak mekanizması çözümü (Alptekin ve diğ., 1986; Tan, 1996; Taymaz ve diğ., 1999) ile Bartın depremi, Karadeniz'in güneyindeki aktif bindirmeye ait ilk sismolojik kanıtı verme özelliğine sahiptir. 15 Ekim 2016 tarihinde, İstanbul'un 195 km kuzeydoğusunda, Zonguldak'ın ise 124 km kuzeybatısında kalan ve yüzeyden 11.4 km derinde olan bir merkez üssüne sahip olarak meydana gelen Karadeniz depremi ($M_l$=5.0) (KRDAE, 2016), Karadeniz'in Türkiye kıyılarına yakın bölgede aletsel olarak kaydedilebilen nadir büyük depremlerden biridir. İstanbul, Kocaeli, Düzce, Sakarya, Zonguldak ve hatta Bulgaristan'ın Varna ve Burgaz şehirlerinde hissedilen bu deprem, yerel saatle 11:18'de gerçekleşmiş ve yaklaşık olarak 7-8 saniye sürmüştür. 15 Ekim 2016 Karadeniz Depremi'nin moment tensör çözümünden elde edilen odak mekanizması çözümü (Url-1), 03 Eylül 1968 Bartın depreminin ($m_b$=5.7, $M_S$=6.6) (Alptekin ve diğ., 1986) odak mekanizması çözümüne benzer bir biçimde ters faylanma göstererek, bu depremin de bölgede var olan aktif bindirmeye ait diğer bir sismolojik kanıt olduğunu ortaya koymuştur.

Karadeniz bölgesinde son yıllarda deprem istasyonlarının sayısının artması, deprem algılama eşiğinin de olumlu yönde düşmesine neden olmuştur. Bölgede son yıllarda meydana gelen depremler incelendiğinde, hemen hemen tüm depremlerin 03 Eylül 1968 Bartın depreminin faylanma türü ile aynı ve/veya baskın ters faylanma bileşenlerine sahip olarak gerçekleştiğini ortaya koymaktadır (Kalafat ve Toksöz, 2015). Bu depremler bölgenin asismik olmayıp, çok sık aralıklarla olmasa da zaman zaman deprem ürettiğini ortaya koymaktadır. Meydana gelen depremlerin odak derinliklerinin genel olarak 10-35 km aralığında değiştiği görülmektedir. Karadeniz'de özellikle son 6 yılda meydana gelen depremlerin fay düzlemi çözümlerine bakıldığında, Güney Karadeniz'de ve orta bölümde özellikle Bartın açıkları, Kastamonu civarı-Samsun açıklarında oblik faylanma karakterli ve sıkışmalı bir tektonik rejimi ifade eden ters fay bileşeni ağırlıklı depremlerin meydana geldiği görülmektedir. Ters faylanma bileşeninin güçlü olduğu depremler genelde havzaya paralel olarak uzanmaktadır. Bu fay geometrisi de sıkışmalı bir gerilmenin bölgede etkili olduğunu göstermektedir (Kalafat, 2018).

Karadeniz'de $M_w \leq 4.0$ olarak gerçekleşen depremler de dikkate değer sayıdadır. Karadeniz, sınırlarına doğru artış gösteren bir depremselliğe sahiptir ve en büyük depremleri sınırlarında gerçekleşmektedir. Odak mekanizmaları, bir miktar D-B bileşenine sahip olan, ağırlıklı olarak K-G yönlü sıkışmaya işaret etmektedir ve bu mekanizmalar levha hareketleri ile de uyumludur. Arap Levhası'nın hareketinin etkisinin büyük bir kısmını Anadolu Levhası'nın batıya doğru olan hareketi ve Kafkasların K-G deformasyonu karşılarken, yalnızca ufak hareketler (yılda yaklaşık 1 mm) Pontidlere doğru aktarılmakta ve Karadeniz K-G yönlü olarak sıkıştırılmaktadır (Kalafat, 2018).

## DENİZ YANSIMA SİSMİĞİ VERİLERİ VE VERİ-İŞLEM AŞAMALARI

17-29 Eylül 1998 tarihleri arasında İstanbul Teknik Üniversitesi (İTÜ), Cambridge Üniversitesi ve Maden Tetkik ve Arama Enstitüsü (MTA) işbirliğiyle, Batı Karadeniz Havzası'nın güneyinde, Akçakoca-Cide arasında kalan bölgenin açıklarındaki şelf ve yamaç bölgesi üzerinde, R/V MTA Sismik-1 gemisi ile 2B sismik yansıma çalışması gerçekleştirilmiştir. Çalışmada kullanılan kaynak ve alıcıların özellikleri Tablo 1.'de verilmektedir. Tablo 2.'de ise çalışmada 14 hat üzerinde toplanan sismik verilere ait veri toplama parametreleri görülmektedir. Sismik verilerin toplandığı, toplam uzunlukları 460 km'yi bulmakta olan 14 hattın konumları Şekil 4'te verilmektedir.

**Tablo 1**. R/V MTA Sismik-1 gemisi ile gerçekleştirilen sismik yansıma çalışmasında kullanılan kaynak ve alıcıların özellikleri.

**Table 1.** The properties of the source and the receiver used at the seismic reflection study performed by R/V MTA Sismik-1.

| **Kaynak (Hava Tabancası, Air Gun)** | |
|---|---|
| Model | Generator-Injector (GI) |
| Adet | 10 |
| Basınç (psi) | 1500 |
| Toplam hacim ($in^3$) | 1380 |
| Derinlik (m) | 5 |
| **Alıcı (Alıcı Kablosu, Streamer)** | |
| Uzunluk (m) | 900 * |
| Derinlik (m) | 10 |
| Hidrofon grubundaki hidrofon sayısı | 16 |
| Hidrofon grupları aralığı (m) | 12.5 |
| Hidrofon grup sayısı | 12 |

* Yalnızca 01 numaralı hattın streamer uzunluğu 1050 m dir.

**Tablo 2.** 14 hat üzerinde toplanan sismik yansıma verilerine ait veri toplama parametreleri.

**Table 2.** Data acquisition parameters of the seismic reflection data collected on 14 lines.

| **Parametreler** | |
|---|---|
| Atış aralığı (m) | 50 |
| Ofset aralığı (m) | 150 |
| Grup aralığı (m) | 12.5 |
| Örnekleme aralığı (ms) | 2 |
| Kanal sayısı | 72 * |
| Katlama (fold) sayısı | 9 * |
| Kayıt uzunluğu (ms) | 8192 |

* Yalnızca 01 numaralı hattın kanal sayısı 84 tür ve katlama sayısı 10'dur.

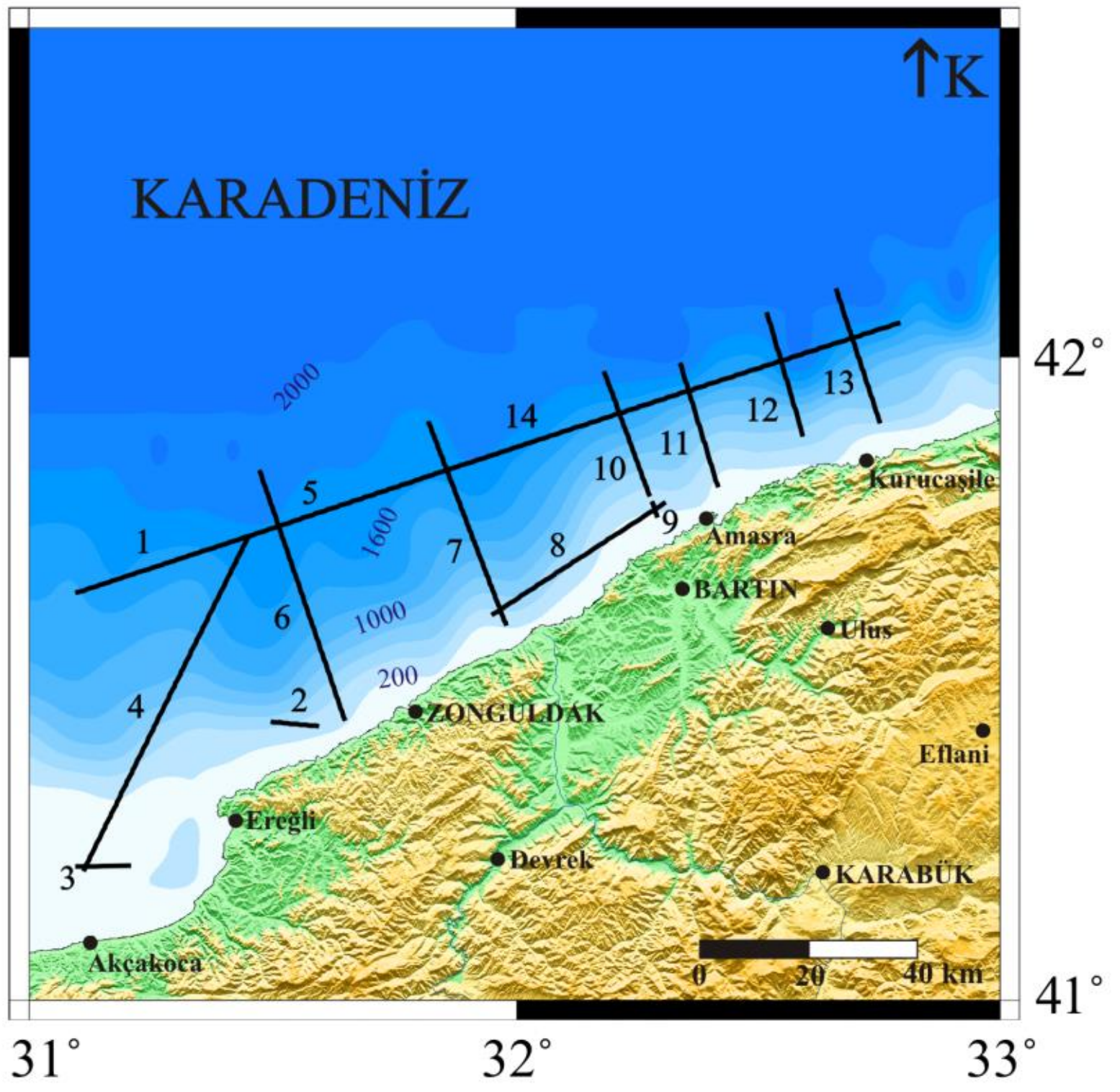

**Şekil 4.** Batı Karadeniz havzasının güneyinde, Akçakoca-Cide arasında kalan bölgenin açıklarında gerçekleştirilen çalışmadaki 14 sismik hattın konumları (Batimetri için GEBCO (IOC, IHO ve BODC, 2003) verileri ve topoğrafya için USGS'in GTOPO-30 küresel verileri ile GMT (Wessel ve diğ., 2013) kullanılarak hazırlanmıştır).

**Figure 4.** The locations of 14 seismic lines of the study performed offshore the southern part of the Western Black Sea Basin (between Akçakoca and Cide) (prepared by using GMT (Wessel et al., 2013) bathymetry data from GEBCO (IOC, IHO and BODC, 2003) and topography data from USGS (GTOPO-30 Global Topography Data)).

Veri toplama aşamasında elde edilen ham verilerin yeraltındaki jeolojik yapıların tanımlanmasında kullanılabilecek yeterli görsel çözünürlüğe sahip sismik kesitlere dönüştürülmesi, bir dizi sismik veri-işlem aşaması sonucu mümkün olmaktadır. Çalışmada toplanan sismik yansıma verilerine, İstanbul Teknik Üniversitesi Jeofizik Mühendisliği Bölümü "Nezihi Canıtez Veri-İşlem Laboratuvarı" bünyesinde bulunan, Paradigm® firması tarafından geliştirilmiş, Unix® işletim sistemi altında çalışan Disco/Focus® ile Redhat® Enterprise Linux® işletim sistemi altında çalışan Echos® sismik veri-işlem paketleri kullanılarak, Şekil 5'te verilen akış diyagramındaki veri-işlem aşamaları sırasıyla uygulanmıştır.

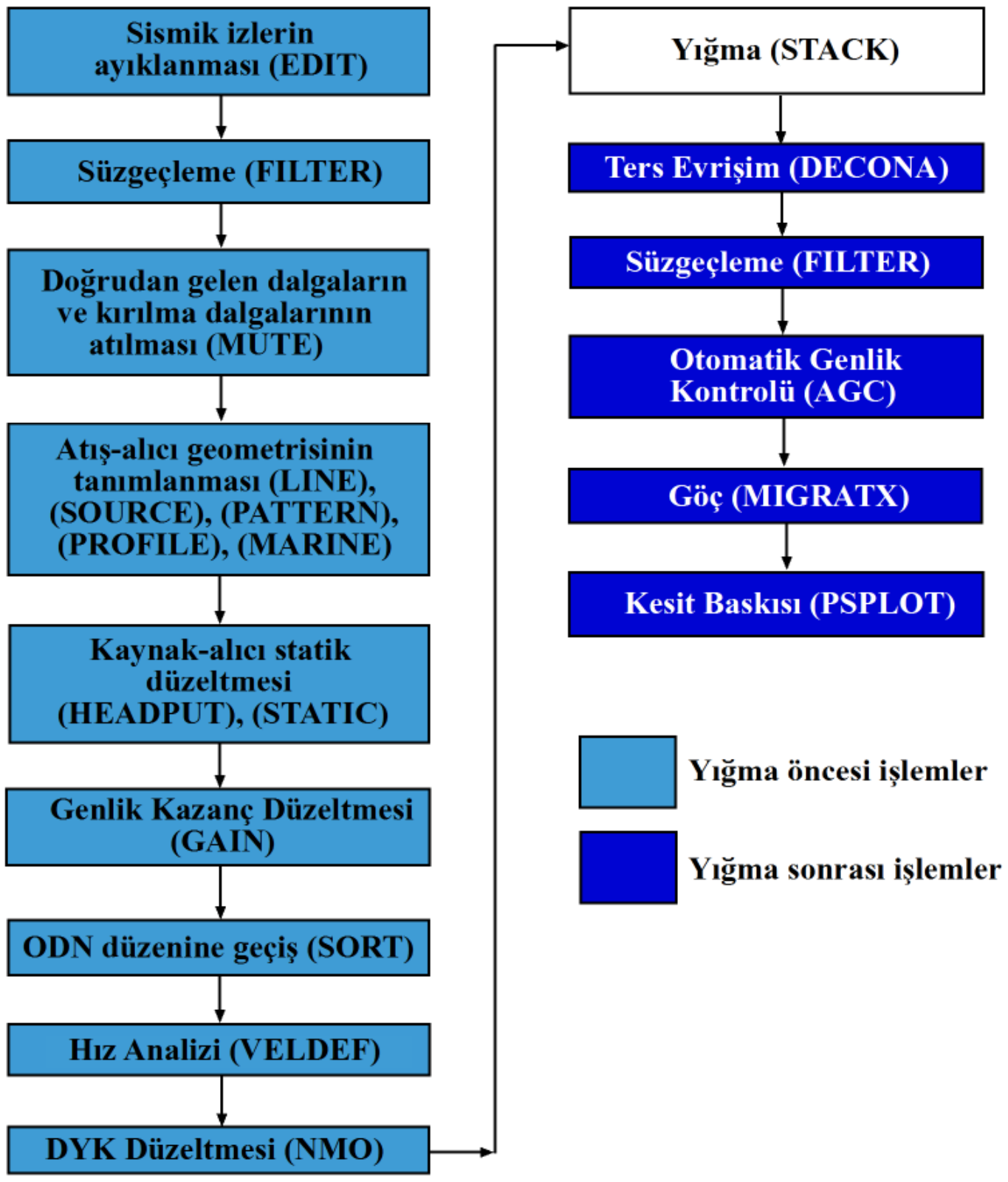

**Şekil 5.** Sismik yansıma verilerine uygulanan veri-işlem aşamalarını ve uygulanan aşama ile ilişkili olarak kullanılan Disco/Focus® modüllerini gösteren akış diyagramı.

**Figure 5.** The flow diagram showing data processing stages and the related modules of Disco/Focus® for seismic reflection data.

## KUYU VERİLERİ

Bu çalışmada, karadaki jeolojinin deniz altında nasıl devam ettiğine dair bilgilerin ortaya konulması amacıyla, çalışma alanı içerisinde karada ve denizde açılmış olan 9 kuyuya ait veriler, Petrol İşleri Genel Müdürlüğü (PİGM)'nden temin edilmiştir. Konumları Şekil 6'da verilen Akçakoca-1, Akçakoca-2, Ereğli-1, Filyos-1, Bartin-1, Ulus-1, Amasra-1, Çakraz-1 ve Gegendere-1 kuyularına ait kompozit kuyu loglarından elde edilen veriler, konumları Şekil 4'te verilen 14 hat üzerinde toplanan sismik veriler ile bir arada değerlendirilerek jeolojik kesitler oluşturulmuştur. Denizde açılmış olan kuyulardan Akçakoca-1 kuyusuna ait kompozit logun (Ünal ve Coşkun, 1976) sadeleştirilmiş hali ve Akçakoca-2 kuyusuna ait kompozit logun (Ünal, 1976) sadeleştirilmiş hali Şekil 7'de verilmektedir. Gazlı kuyu olarak terkedilen Akçakoca-1 kuyusunun son derinlikte ulaştığı formasyon Kretase yaşlı Hamsaros formasyonudur (Ünal ve Coşkun, 1976). Gaz emareli kuyu olarak terk edilen Akçakoca-2 kuyusunun ise son derinlikte ulaştığı formasyon Eosen yaşlı formasyondur (Ünal, 1976).

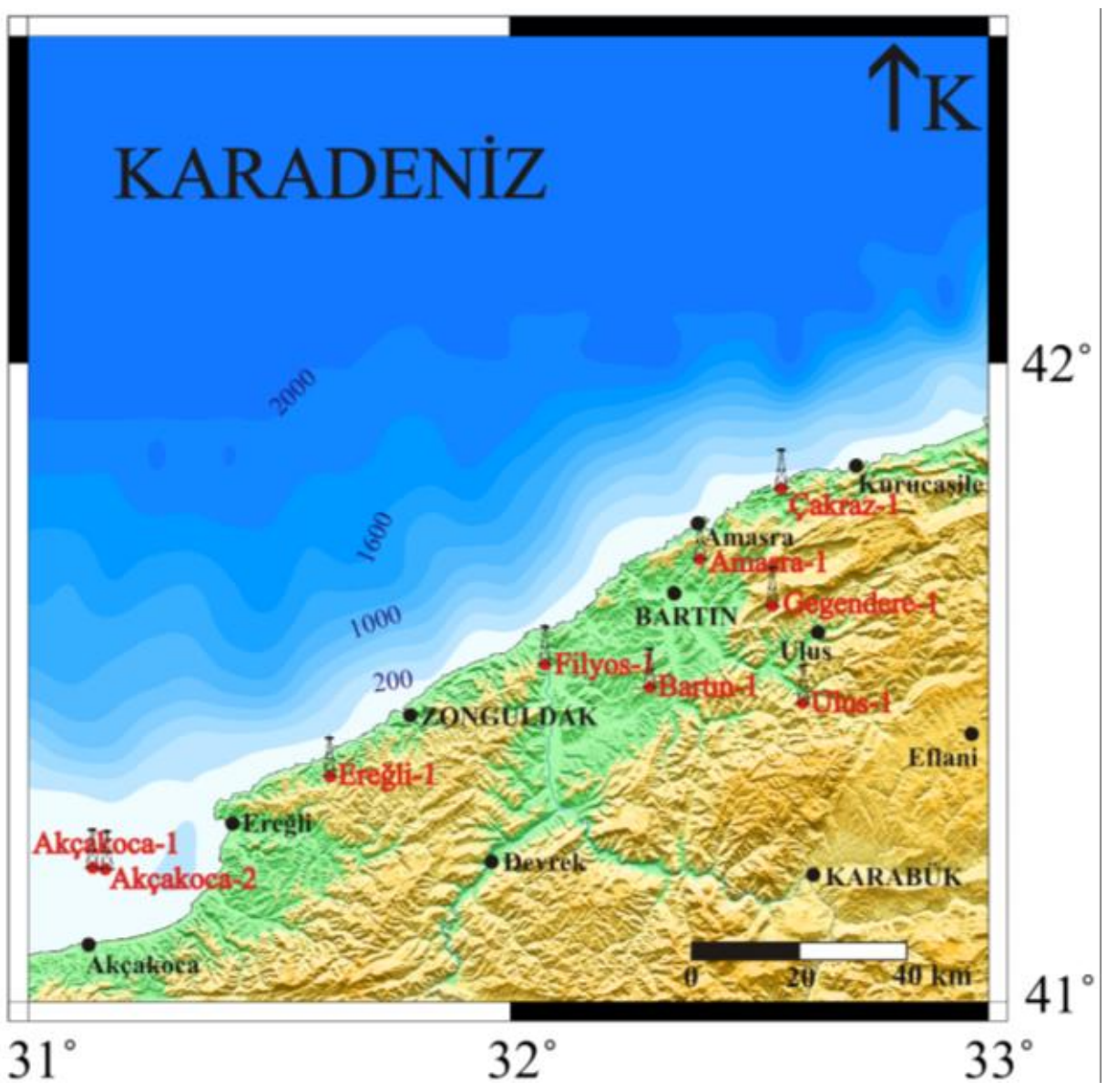

**Şekil 6.** Çalışma alanı içerisinde bulunan dokuz araştırma kuyusunun konumları (GMT (Wessel ve diğ., 2013) kullanılarak hazırlanmıştır).

**Figure 6.** The locations of nine exploratory wells in the study area (prepared by using GMT (Wessel et al., 2013)).

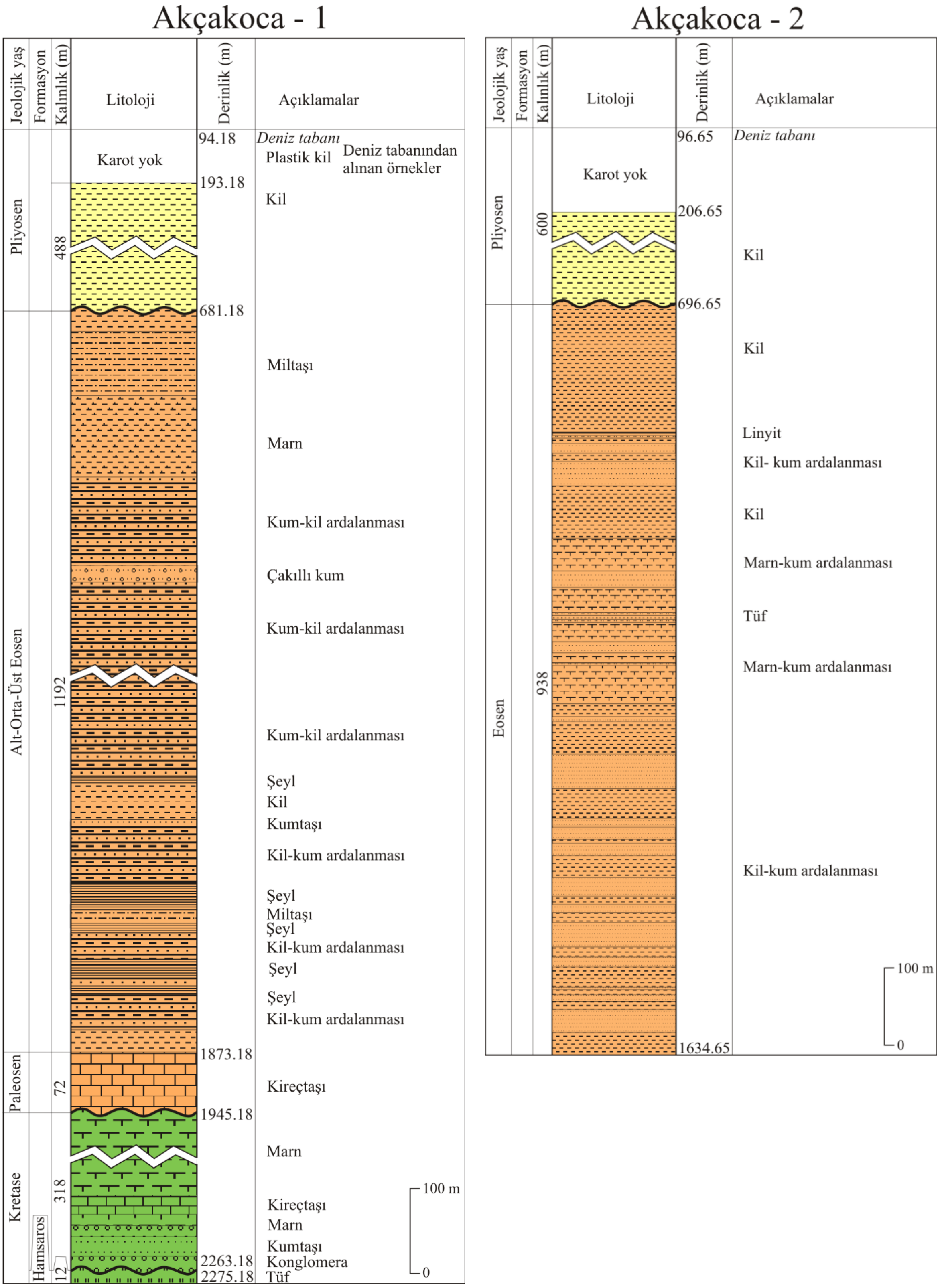

**Şekil 7.** Akçakoca-1 kuyusuna ait sadeleştirilmiş kompozit kuyu logu (Ünal ve Coşkun, 1976) ve Akçakoca-2 kuyusuna ait sadeleştirilmiş kompozit kuyu logu (Ünal, 1976).

**Figure 7.** The simplified composite well logs of Akçakoca-1 well (Ünal and Coşkun, 1976) and Akçakoca-2 well (Ünal, 1976).

## YORUMLAMA

Dağ kuşaklarının oluşumunda çok önemli bir rol oynayan bindirme fayları (thrust faults), dev kaya kütlelerinin yükselimine ve ötelenmelerine sebep olmaktadırlar. Tavan bloğun yukarıya doğru taban blok üzerine hareket ettiği eğim atımlı faylar olan ters faylar, eğim açısı 30°'den düşük olduğunda bindirme fayları adını almaktadır. Anderson (1951), artık klasikleşmiş olan çalışmasında, her üçü de farklı büyüklüklere sahip olan (en büyük ($\sigma_1$), orta ($\sigma_2$) ve en küçük ($\sigma_3$)) üç ana gerilme eksenini tanımlamaktadır. Bu üç ana gerilme ekseninin her birinin düşey konumda olması durumuna göre, üç farklı tektonik rejim ortaya koymuştur. $\sigma_1$'in düşey olduğu durum normal-fay rejimi, $\sigma_2$'nin düşey olduğu durum ise doğrultu-atımlı fay rejimi olarak adlandırılmaktadır. Bu rejimlerden üçüncüsü olan ve ters fayların oluşumuna olanak sağlayan bindirme-fayı rejiminde (thrust-fault regime) (Şekil 8a), en küçük olan $\sigma_3$ ana gerilmesinin düşeyde olduğu ve diğer iki ana gerilmenin de ($\sigma_1$ ve $\sigma_2$) yatayda olduğunu belirtmektedir (Fossen, 2010).

Kıvrım kuşakları için özellikle önemli olan bir gerçek şudur ki, kıvrımlara genellikle ters faylar eşlik etmektedir. Bu ters fayların büyük bir kısmı düşük açılı olan bindirme faylarıdır ve neticede oluşturdukları bu yapısal alanlarda, kıvrım-ve-bindirme kuşağı (fold-and-thrust belt) olarak adlandırılmaktadır. Bazı bindirme fayları ve yüksek açılı ters faylar yüzey kırığı oluşturabilirler, fakat gömülü ters faylar (buried reverse faults) olarak adlandırılan diğer birçok ters fay ise antiklinallerin çekirdeklerinde saklı olarak kalırlar (Keller ve Pinter, 2002). Son zamanlarda bu gömülü aktif fayların belirgin şekilde deprem tehlikesi ortaya koydukları kabul edilmektedir ki, bu büyük zararlara sebep olan depremler, kıvrımlı kayaçların içinde veya tam altında konumlanmış olan faylar üzerinde oluşabilmektedir. Bu özellikteki faylar, kilometrelerce derinlikte gömülü olabilirler ve deprem sırasında yırtıldıklarında yüzeyde yükselme (uplift) ve kıvrımlanmalara sebep olabilirler (Stein ve Yeats, 1989; Keller ve Pinter, 2002).

Faylanma, kıvrımlanma ve kıvrım-ve-bindirme kuşakları arasındaki ilişkileri açıklayabilmek amacıyla, gömülü fayların oluşumuyla sonuçlanan kıvrımlanma davranışlarının öngörülebilmesi için birçok farklı model geliştirilmiştir. Bu geliştirilen kıvrım modellerinden bir tanesi fay-ilerleme kıvrımları (fault-propagation folds) olarak adlandırılan modeldir (Keller ve Pinter, 2002). Fay-ilerleme kıvrımları, ilerleyen bir fayın önünde oluşturduğu deformasyondan kaynaklanmaktadır. Bir bindirme fayının ucundaki fay-ilerleme kıvrımlanması aşamaları Şekil 8b'de görülmektedir.

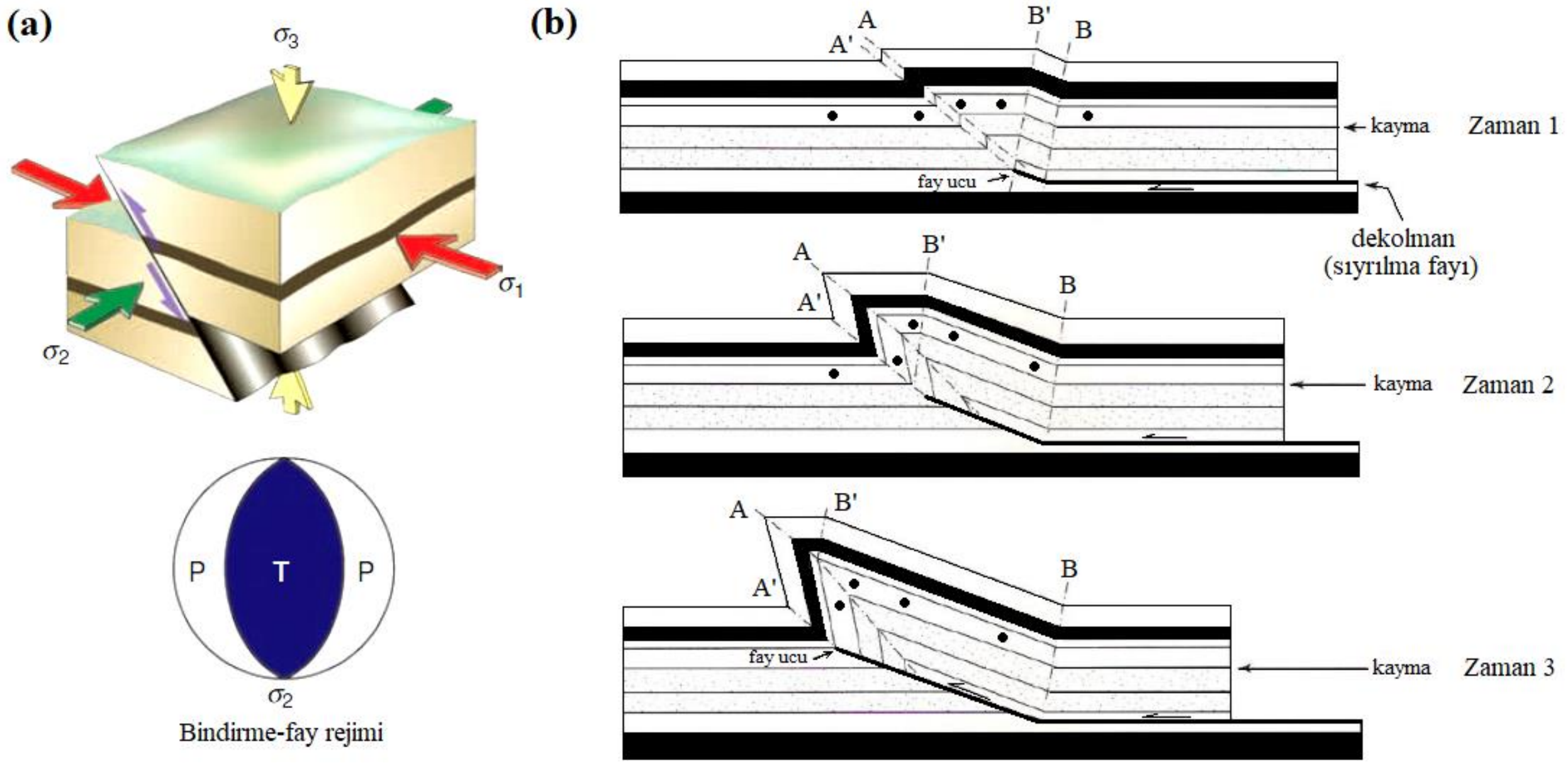

**Şekil 8. (a)** Anderson (1951)'a göre ana gerilmelerin ($\sigma_1$, $\sigma_2$, $\sigma_3$) yönelimleri ile bindirme-fay tektonik rejiminin arasındaki ilişki. Fay düzlemi çözümü sıkışma (P) ve gerilme (T) alanlarını göstermektedir (Fossen, 2010). **(b)** Bir dekolman (sıyrılma fayı) boyunca kayma sonucunda bir fay-ilerleme kıvrımının ve daha dik bir bindirme fayının gelişimi (Suppe, 1985).

**Figure 8. (a)** Relationship between the orientation of principal stresses and thrust-fault tectonic regime according to Anderson (1951). Fault plane solution shows fields of compression (P) and tension (T) (Fossen, 2010). **(b)** Development of a fault-propagation fold as a result of slip along a décollement (detachment fault) and a steeper thrust fault (Suppe, 1985).

İlerleyen fayın ucu, zayıf bir stratigrafik seviye boyunca gelişen sıyrılma fayından (detachment fault, décollement) başlayarak fay bükülme kıvrımları boyunca ilerleyişine devam eder. Sıyrılma fayı altındaki kayaçlar ile üzerindeki kayaçlar karşılaştırıldıklarında alttaki kayaçların genellikle deformasyona uğramadığı görülür. Kıvrım geliştikçe daha asimetrik bir özellik kazanır. Bindirme fayı yaygın olarak kıvrımın içinde son bulmaktadır (Suppe, 1985; Keller ve Pinter, 2002). Ayrıca, fay ve kıvrımdaki yatay ilerleme oranı, düşey kayma oranından birkaç kez daha büyük olabilmektedir (Keller ve diğ, 1998; Keller ve diğ., 1999; Keller ve Pinter, 2002).

Dilimli (imbricate) bindirme ve dupleks bindirme (duplexes) gibi bindirme fay sistemleri, kıvrım-ve-bindirme kuşakları için önemli bileşenler olarak kabul edilmektedir. Bir dilimli bindirme sistemi, aralarında belirli uzaklıklar bulunan çok sayıda bindirme fayından meydana gelmektedir. Fay-ilerleme kıvrımları sisteminden gelişerek oluşan bu dilimli fay sistemlerinde, bindirme fayları arası uzaklık fazla ise, birbirini takip ederek oluşan kıvrımlar, ilk bindirmenin sahip olduğu geometrinin benzeri biçimde oluşmakta ve böylece, daha sonra oluşan bindirmelerin geometrisinde herhangi bir değişime sebep olmamaktadır. Bindirme fayları arası uzaklık az ise, ilk kıvrımdaki bindirmeyle karşılaştırıldığında, gelişen sonraki kıvrımlarda dereceli olarak daha dik bindirmeler oluşmakta ve bu bindirmelere bağlı olarak oluşan kıvrımların geometrilerinde de, gittikçe artış gösteren dönme hareketi sebebiyle, farklılıklar gözlenmektedir (Mitra, 1986). Bu çalışmada, kıyıya dik olarak uzanan hatlara ait oluşturulan zaman ortamı sismik göç kesitlerinden, aralarındaki uzaklıklar az olan birçok bindirme fayından oluşmuş dilimli fay sistemlerini temsil eden yapılara görsel örnekler vermek mümkün olmaktadır.

Sismik hatlardan 3 numaralı hatta ait veriler, veri toplama aşamasında Akçakoca-1 ve Akçakoca-2 kuyularının bulunduğu konumlar üzerinden seyir edilerek toplanmıştır (Şekil 9a). Bu nedenle, 3 numaralı hattın zaman ortamı sismik göç kesitinde ayırt edilen sismik birimlerin jeolojik yaşlandırmaları doğrudan, Şekil 7'de verilen Akçakoca-1 ve Akçakoca-2 kuyularının kompozit loglarından faydalanılarak gerçekleştirilmiştir. Diğer hatlarda ayırt edilen sismik birimlerin jeolojik yaşlandırmaları içinde yine bu kompozit loglardan yararlanılmıştır. Şekil 9b'de, 3 numaralı hatta ait sismik göç kesiti verilmiştir. Şekil 9c'de ise 3 numaralı hatta ait sismik göç kesitinin, Akçakoca-1 ve Akçakoca-2 kuyularının kompozit kuyu loglarından faydalanılarak jeolojik olarak yorumlanmış hali görülmektedir. Sismik kesitte özellikle iki kuyu arasında kalan bölge incelendiğinde, Eosen birimlerinin üst sınırından başlayarak, yüzeye doğru dike yakın bir dalım açısına sahip ve batıdan doğuya doğru derinlere inildikçe dalım açısında azalma görülen bir fay ayırt edilmiştir. Bu fayın, bölgedeki sıkışmalı tektonik rejimin etkisiyle oluşmuş olması muhtemeldir.

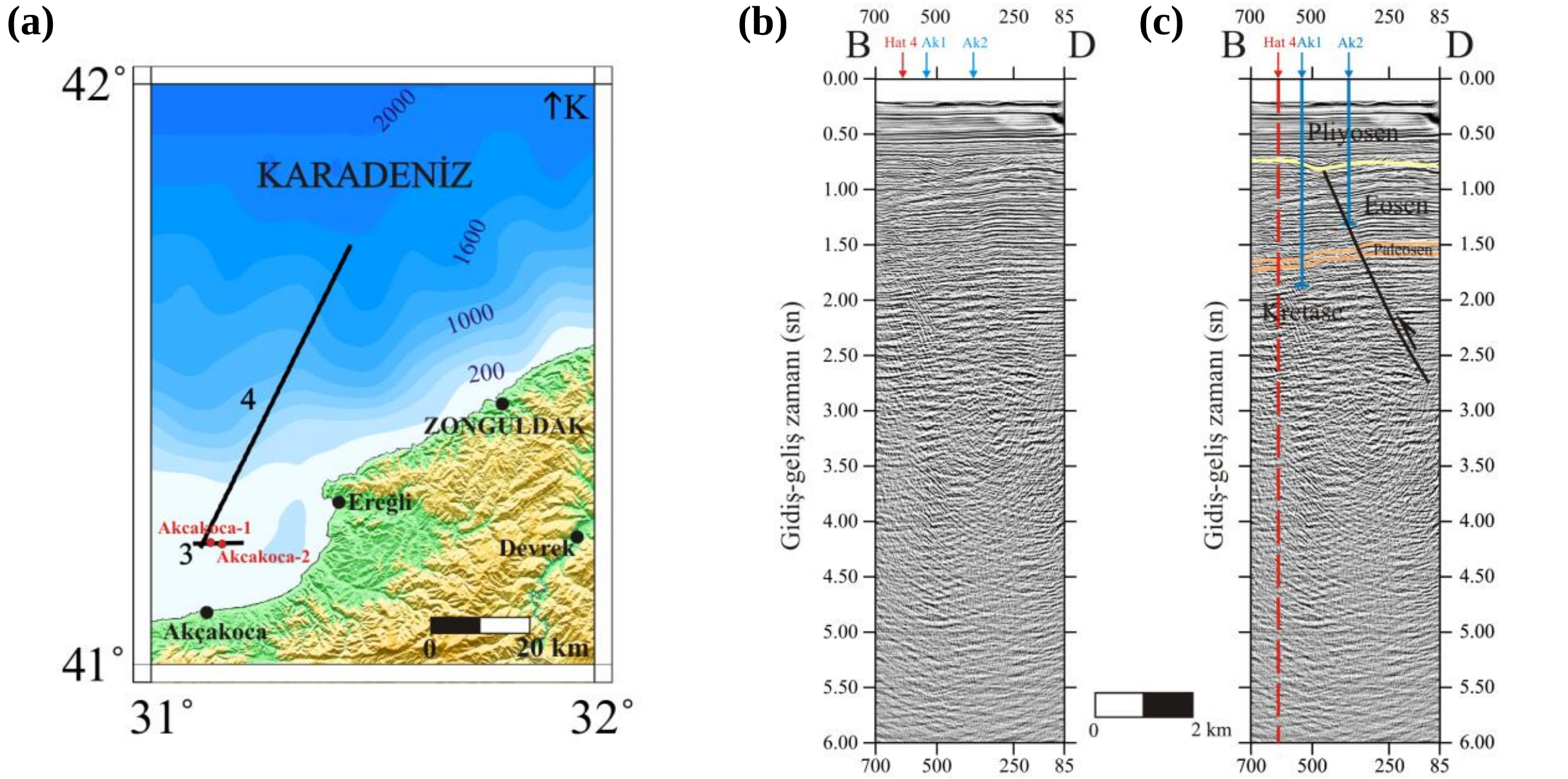

**Şekil 9. (a)** Akçakoca-1, Akçakoca-2 kuyuları ve 3 ve 4 numaralı sismik hatların konum haritası. **(b)** 3 numaralı hatta ait sismik göç kesiti (2 × Düşey Büyütme). **(c)** 3 numaralı hatta ait sismik göç kesitinin, Akçakoca-1 ve Akçakoca-2 kuyularının kompozit kuyu loglarından faydalanılarak jeolojik olarak yorumlanmış hali.

**Figure 9. (a)** The location map of Akçakoca-1, Akçakoca-2 wells and seismic Lines 3 and 4. **(b)** Seismic migration section of Line 3 (2 × Vertical Exaggeration). **(c)** Geologically interpreted seismic migration section of Line 3 by referencing composite well logs of Akçakoca-1 and Akçakoca-2 wells.

Şekil 10'da 4 numaralı hatta ait sismik göç kesitinin 85 ile 4000 numaralı Ortak Derinlik Noktaları (ODN) arasında kalan bölümü görülmektedir. 3 numaralı hattı kesen bu hattın GB bölümünde, sıkışmalı tektonik rejimin etkisiyle oluşmuş olduğu düşünülen bir fay ayırt edilmiştir. Bu fay, Robinson ve diğ. (1995) ve Robinson ve diğ. (1996)'nin çalışmalarında verilen BP91-202 numaralı sismik hatta ait kesitlerde de gözlemlenmektedir.

Bu çalışmada, jeolojik yaşlandırmaları yapılan tüm sismik kesitlerin, karadaki jeoloji ile ilişkilendirilebilmeleri için, karada gerçekleştirilmiş olan farklı önceki jeolojik çalışmalarda sunulan jeolojik kesitlerden faydalanılmıştır. Buna bağlı olarak, 4 numaralı hattın sismik kesitinden yararlanarak oluşturulan jeolojik kesitle (Şekil 11a), konum haritası Şekil 11b'de görülen, Aydın ve diğ. (1987) tarafından verilmiş jeolojik kesit (Şekil 11c) ilişkilendirilerek, Şekil 12'de verilen ve karadaki jeolojinin deniz altında nasıl devam ettiğine dair bilgilerin sunulduğu jeolojik kesit elde edilmiştir.

Karadeniz kıyısına dik olarak uzanan hatlardan biri olan 11 numaralı hatta ait sismik göç kesiti Şekil 13a'da ve bu hatta ait sismik göç kesitinin jeolojik olarak yorumlanmış hali Şekil 13b'de verilmiştir. Şekil 14a'da, 11 numaralı hatta ait sismik göç kesitinde ayırt edilen, sıkışmalı tektonik rejimin etkisiyle oluştuğu düşünülen fay yapısı görülmektedir. Şekil 14b'de, 11 numaralı hattın sismik kesitinden yararlanarak oluşturulan jeolojik kesit verilmiştir. 11 numaralı hat, 03 Eylül 1968 Bartın depreminin episantırına konum olarak en yakın olan hat olma özelliğindedir. 03 Eylül 1968 Bartın depreminin episantırının konumunun, 11 numaralı hattın ve bu hattın karadaki jeoloji ile ilişkilendirilmesinde yararlanılan Akbaş ve diğ. (2002)'nin sunduğu jeolojik kesitin konumlarının bir arada verildiği harita Şekil 15a'da verilmektedir. Şekil 15b'de ise Akbaş ve diğ. (2002)'nin vermiş oldukları jeolojik kesit görülmektedir. Şekil 14b'deki jeolojik kesit ile Şekil 15b'deki jeolojik kesit kullanılarak oluşturulan kara-kıyı ötesi jeolojik kesiti ise Şekil 16'da verilmektedir.

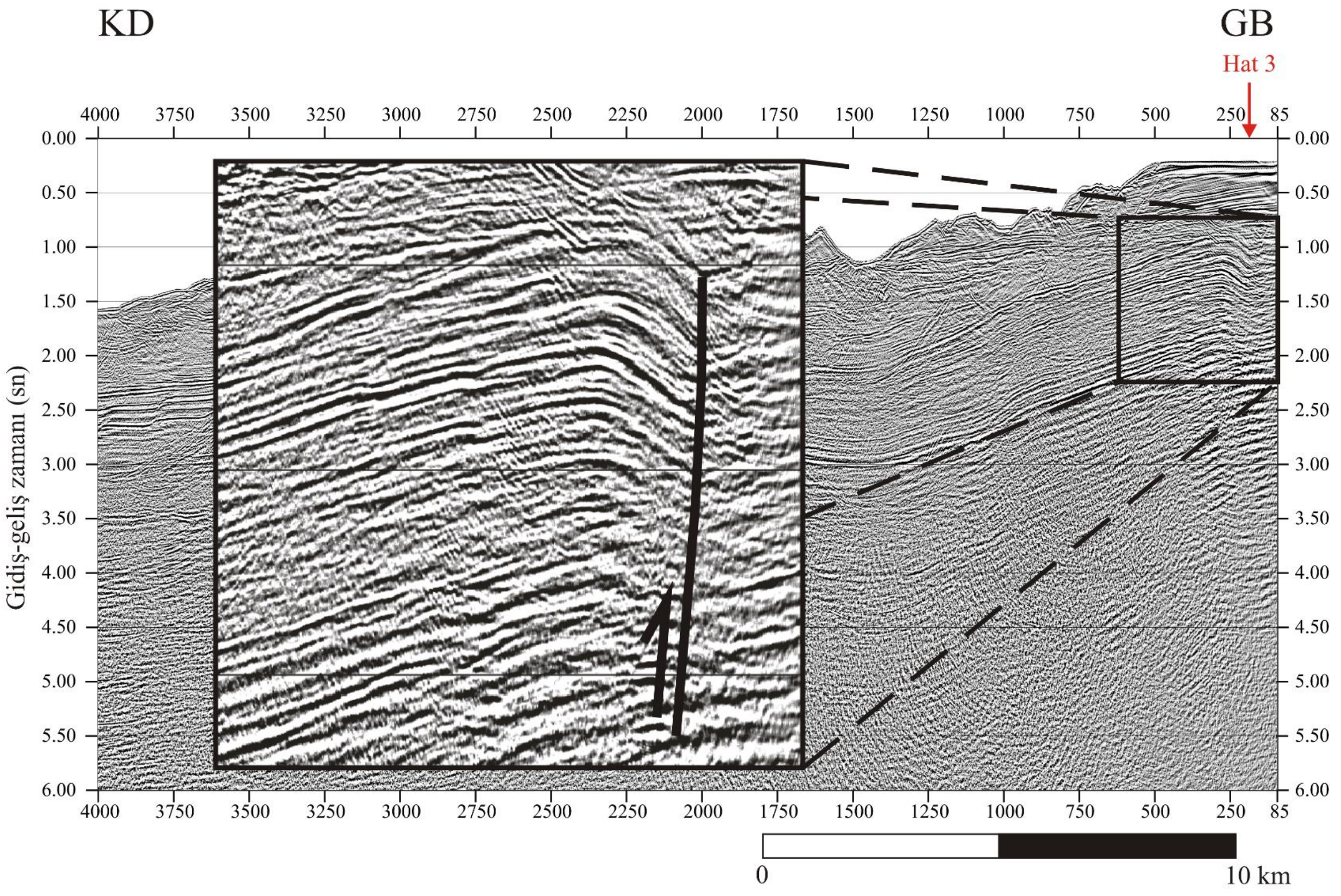

**Şekil 10.** 4 numaralı hatta ait sismik göç kesitinin GB bölümünde ayırt edilen, sıkışmalı tektonik rejimin etkisiyle oluşmuş fay yapısı.

**Figure 10.** The fault structure formed by the effect of the compressional tectonic regime recognized at the SW part of the seismic migration section of Line 4.

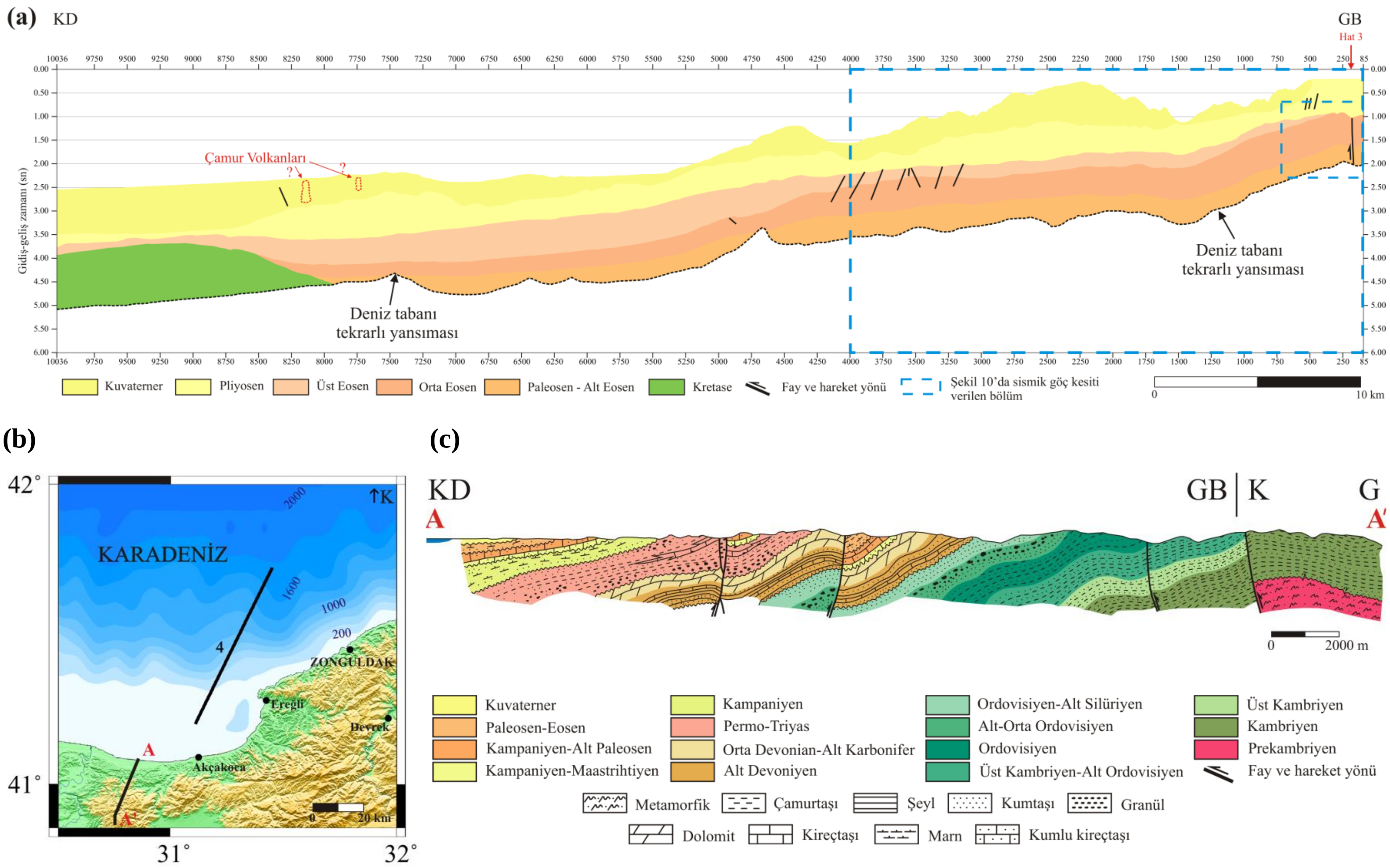

**Şekil 11.** **(a)** 4 numaralı hatta ait sismik göç kesitinden yararlanarak hazırlanan jeolojik kesit (2 ×Düşey Büyütme). **(b)** 4 numaralı hat ve Aydın ve diğ. (1987) tarafından verilen jeolojik kesitin konum haritası. **(c)** Aydın ve diğ. (1987) tarafından verilen jeolojik kesit.

**Figure 11.** **(a)** The geological section prepared by using seismic migration section of Line 4 (2 ×Vertical Exaggeration). **(b)** The location map of Line 4 and the geological section given by Aydın et al. (1987). **(c)** The geological section given by Aydın et al. (1987).

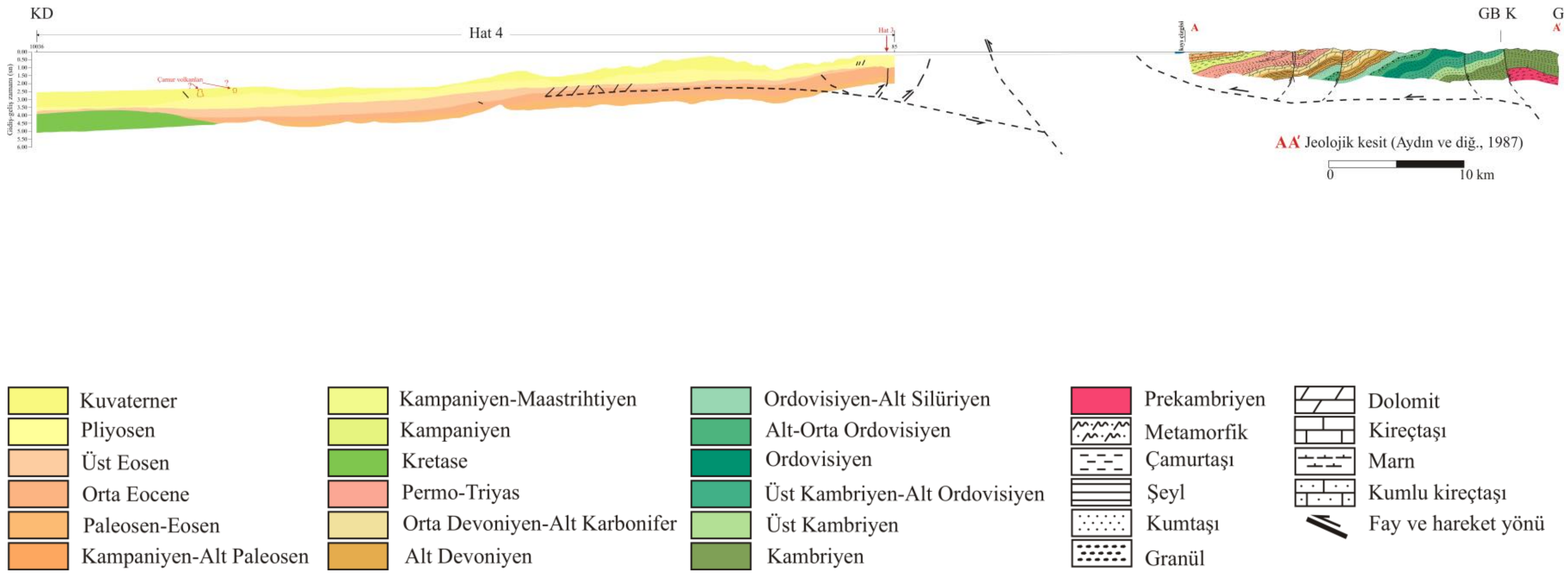

**Şekil 12.** 4 numaralı hatta ait sismik göç kesitinden elde edilen jeolojik kesitin ve Aydın ve diğ. (1987) tarafından verilen jeolojik kesitin kullanılmasıyla yorumsal olarak oluşturulan kara-kıyı ötesi jeolojik kesiti.

**Figure 12.** The interpretively created geological land-offshore transect by using the geological section obtained from seismic migration section of Line 4 and the geological section given by Aydın et al. (1987).

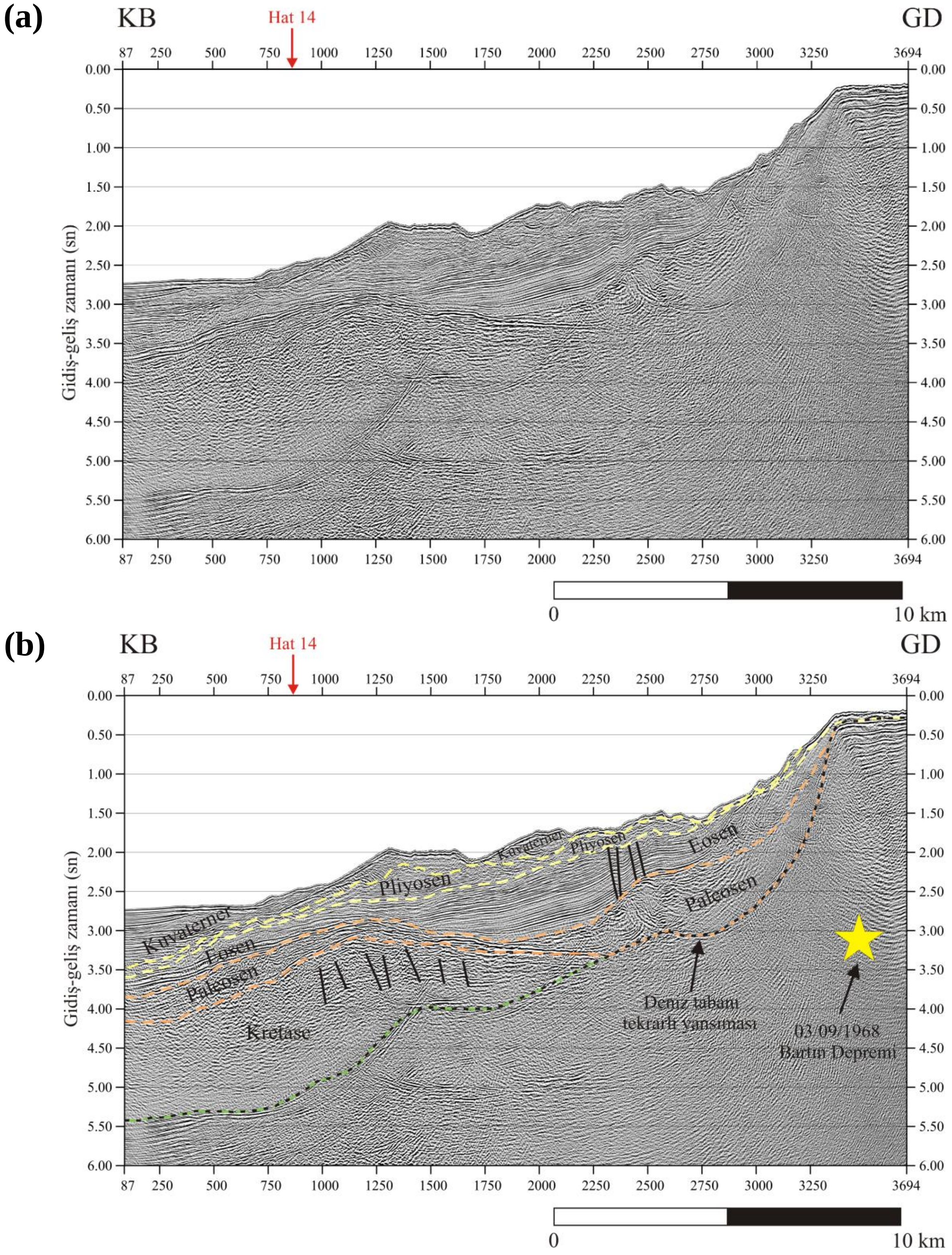

**Şekil 13. (a)** 11 numaralı hatta ait sismik göç kesiti (2 × Düşey Büyütme). **(b)** 11 numaralı hatta ait sismik göç kesitinin jeolojik olarak yorumlanmış hali.

**Figure 13. (a)** Seismic migration section of Line 11 (2 × Vertical Exaggeration). **(b)** The geologically interpreted seismic migration section of Line 11.

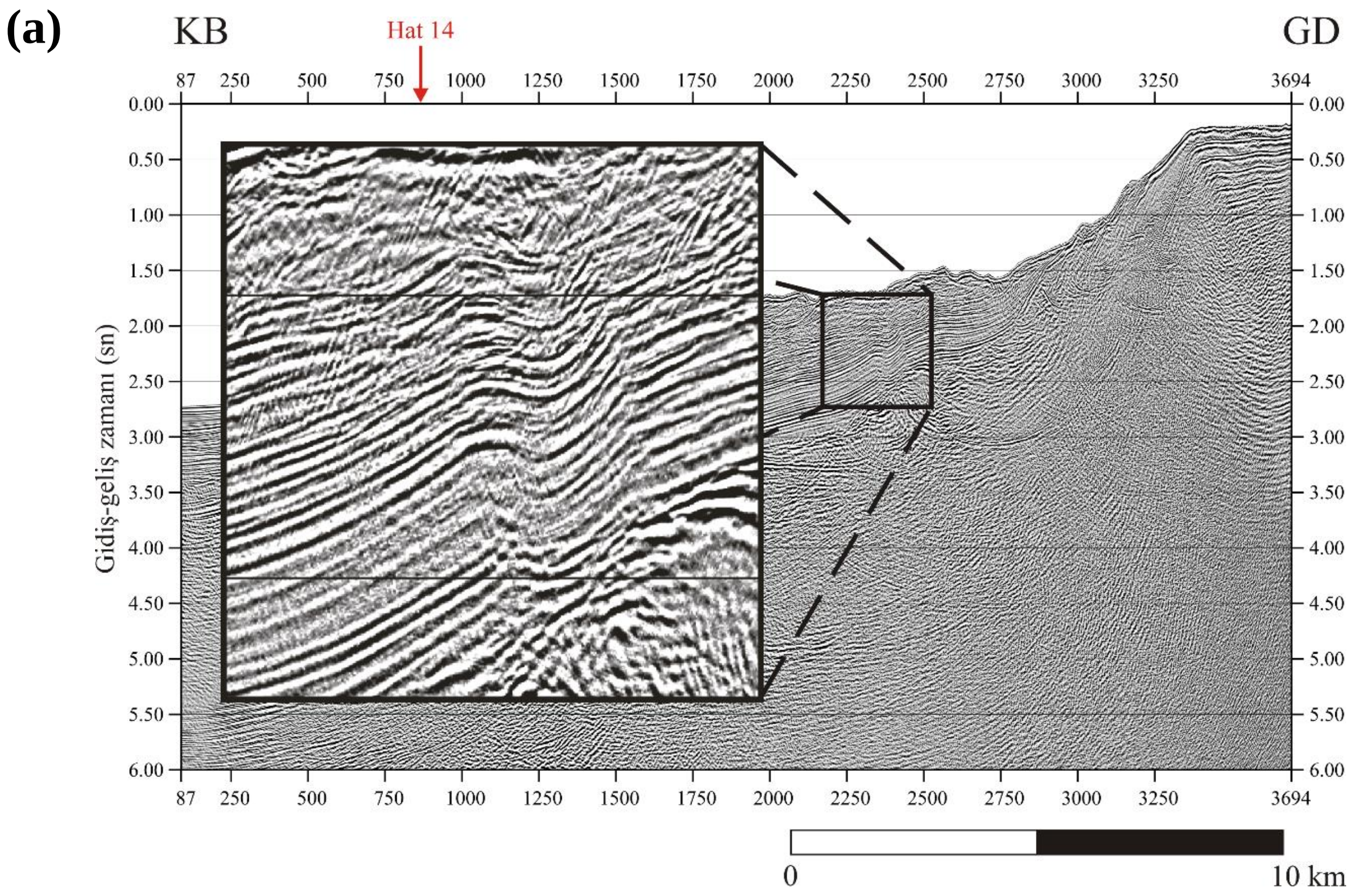

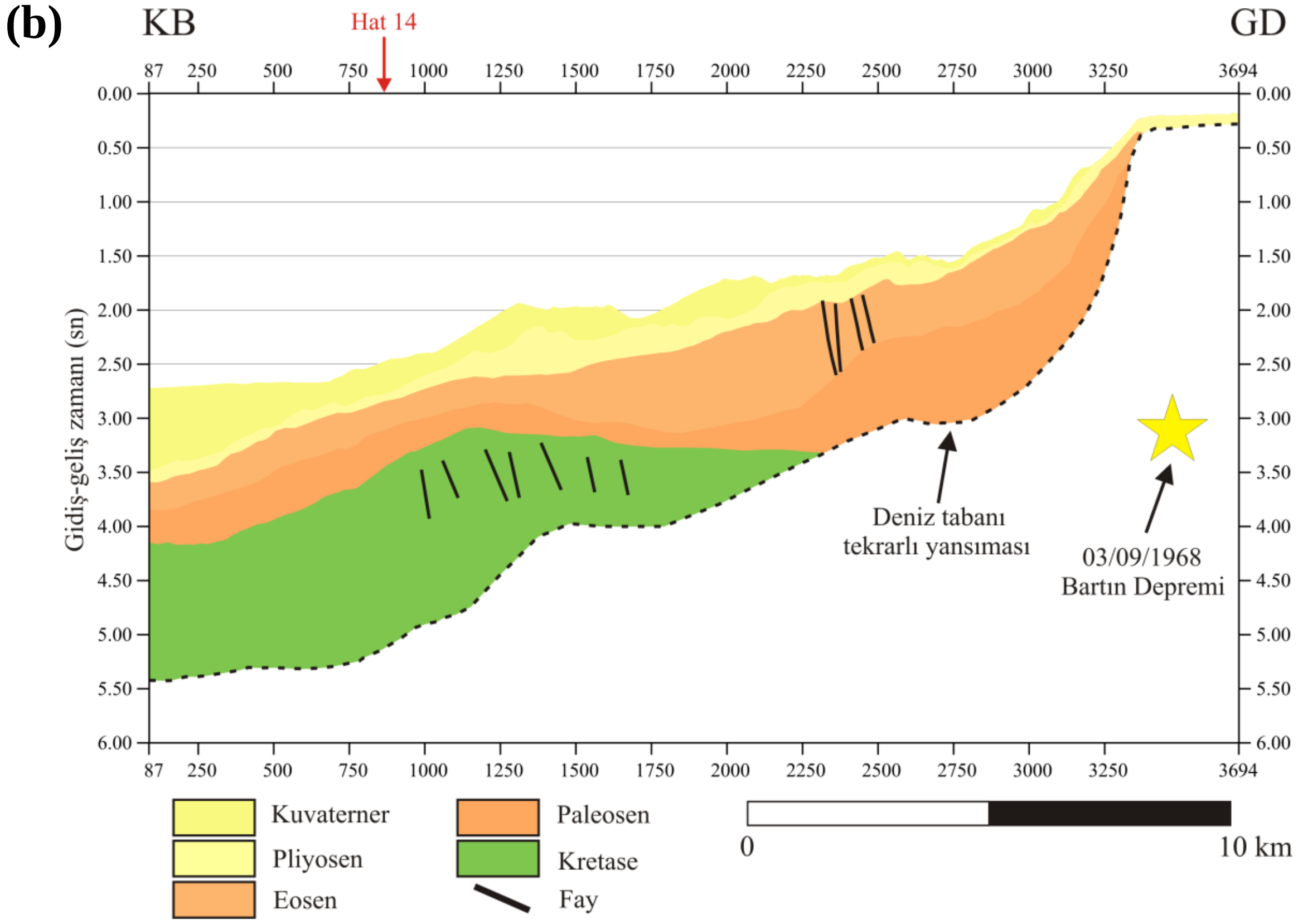

**Şekil 14. (a)** 11 numaralı hatta ait sismik göç kesitinde ayırt edilen, sıkışmalı tektonik rejimin etkisiyle oluşmuş fay yapısı. **(b)** 11 numaralı hatta ait sismik göç kesitinden yararlanarak hazırlanan jeolojik kesit (2 × Düşey Büyütme).

**Figure 14. (a)** The fault structure formed by the effect of the compressional tectonic regime recognized at the seismic migration section of Line 11 **(b)** The geological section prepared by using seismic migration section of Line 11 (2 × Vertical Exaggeration).

(a)

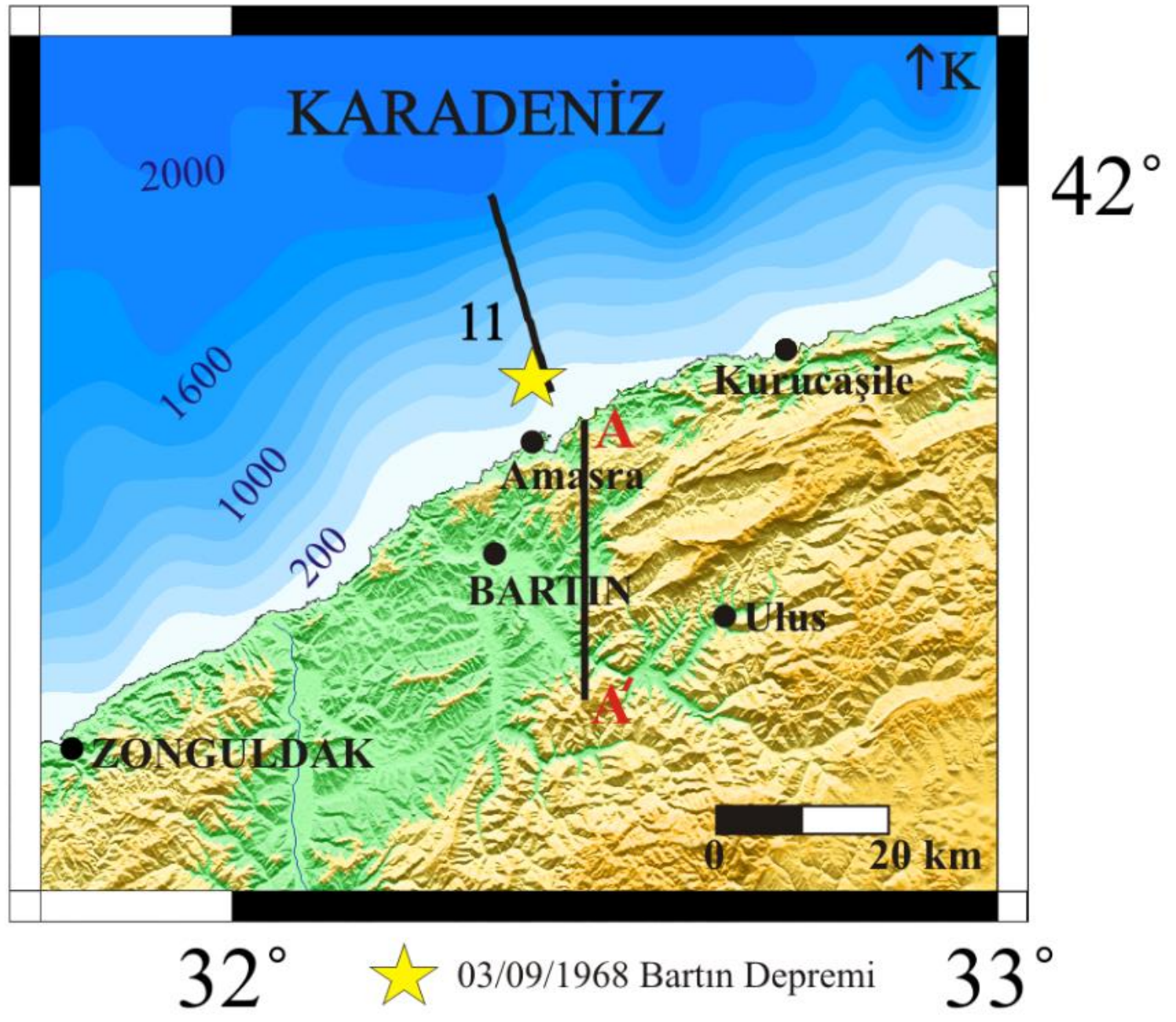

(b)

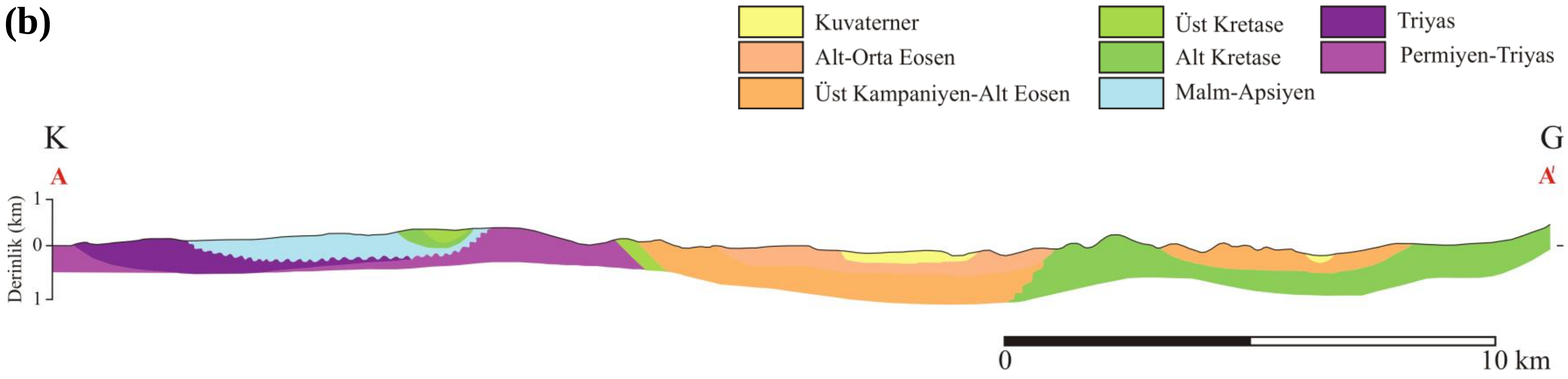

**Şekil 15. (a)** 11 numaralı hat ve Akbaş ve diğ. (2002) tarafından verilen jeolojik kesitin konum haritası. **(b)** Akbaş ve diğ. (2002) tarafından verilen jeolojik kesit.

**Figure 15. (a)** The location map of Line 11 and the geological section given by Akbaş et al. (2002). **(b)** The geological section given by Akbaş et al. (2002).

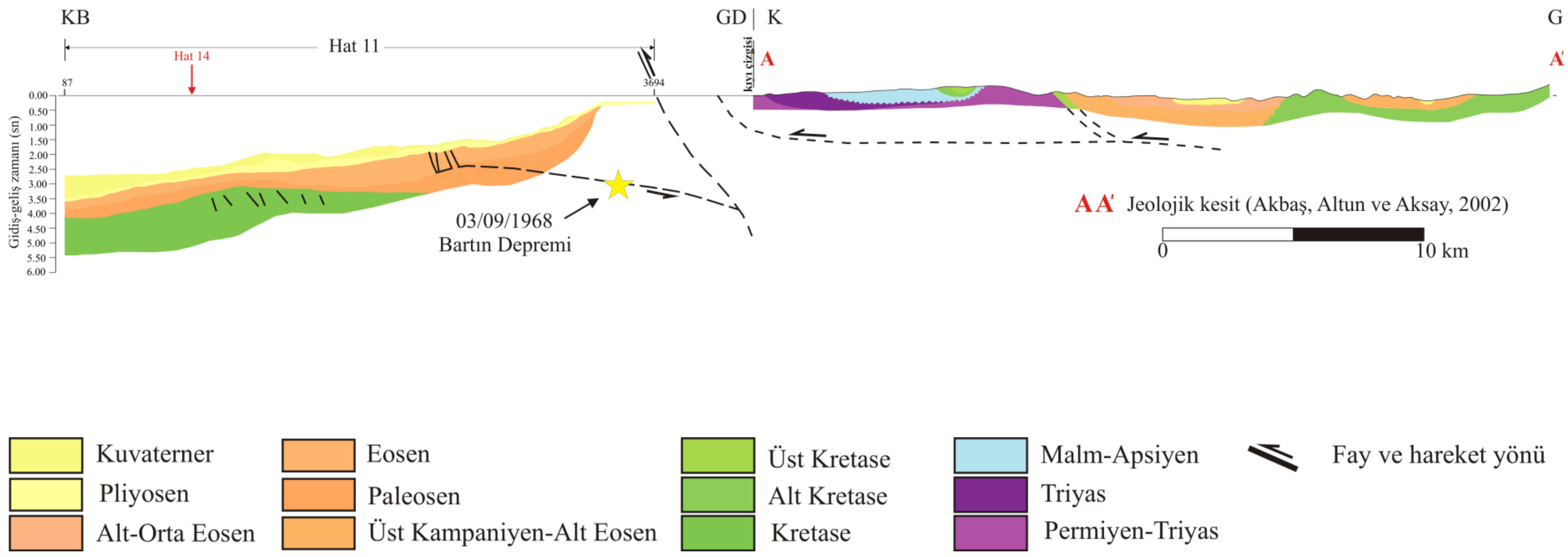

**Şekil 16.** 11 numaralı hatta ait sismik göç kesitinden elde edilen jeolojik kesitin ve Akbaş ve diğ. (2002) tarafından verilen jeolojik kesitin kullanılmasıyla yorumsal olarak oluşturulan kara-kıyı ötesi jeolojik kesiti.

**Figure 16.** The interpretively created geological land-offshore transect by using the geological section obtained from seismic migration section of Line 11 and the geological section given by Akbaş et al. (2002).

## SONUÇLAR

Batı Karadeniz Havzası'nın güneyindeki aktif bindirmeye ait ilk sismolojik kanıt olan 03 Eylül 1968 Bartın depreminden sonra diğer bir kanıt olarak 15 Ekim 2016 Karadeniz Depremi meydana gelmiştir. Bölgedeki sıkışmalı tektonik rejimin etkisiyle oluşan bu deprem, ters faylanma gösteren odak mekanizması çözümüyle, sağ yanal doğrultu atımlı KAF boyunca meydana gelen depremlerden farklılık gösteren depremlerin bölgede meydana geldiğini bir kez daha ortaya koymuştur.

Bu çalışma, Batı Karadeniz Havzası'nın güneyinde yer alan Türkiye kıyılarında, Akçakoca ile Cide arasında kalan bölge açıklarında bulunan şelf ve yamaç bölgeleri altında, sıkışmalı tektonik rejimin etkisi ile oluştuğu düşünülen fayların ve bu fayların aktiviteleriyle oluşan jeolojik yapıların ortaya çıkarıldığı bir ilksel çalışmadır. Çalışmada verilen kara-kıyı ötesi jeolojik kesitleri yardımıyla, çalışma alanı içerisinde kalan bölgede karadaki jeolojinin deniz altında nasıl devam ettiğine dair bilgiler sunulmaktadır.

Nikishin ve diğ. (2015a) çalışmalarında, Batı Pontidlerin, Geç Eosen ve Erken Oligosen zamanlarında kıvrımlandıklarını ve Karadeniz'e doğru bindirdiklerini (Okay ve diğ., 2001; Sunal ve Tüysüz, 2002; Cavazza ve diğ., 2011) vurgulamışlar, bu kıvrım-ve-bindirme kuşağının tabanda bir sıyrılma yüzeyi ile ayrılmakta olduğunu (Sunal ve Tüysüz, 2002) belirtmişlerdir. Çalışmalarında vurguladıkları diğer bir konu ise, Maykop çökelleri tarafından örtülmüş olduğunu ifade ettikleri bu kıvrım ve bindirme yapıları ile sıyrılma yüzeyini, inceledikleri sismik hatlara ait kesitlerde görememiş olmalarıdır. Bu çalışmada, kıyıya dik olarak uzanan hatlara ait oluşturulan zaman ortamı sismik göç kesitlerinden, aralarındaki uzaklıklar az olan birçok bindirme fayından oluşmuş dilimli fay sistemlerini temsil eden yapılara görsel örnekler vermek mümkün olmaktadır. Sismik göç kesitlerinden elde edilen bu görüntüler, çalışma alanını etkileyen sıkışmalı tektonik rejimin varlığının açıklığa kavuşturulması için oldukça faydalı ipuçları vermektedir.

Sismik kesitlerde, gösterdiği farklı yansıma karakterlerinden ötürü, temel kaya ve üzerinde örtülü olan ince ve genç çökellerden oluşan birimler kolaylıkla ayırt edilebilmiştir. Temel kayası olduğu düşünülen Kretase yaşlı birimlerin üzerini örten birimlerin jeolojik yaşlandırmaları, Akçakoca-1 ve Akçakoca-2 kuyularının kompozit logundan faydalanılarak gerçekleştirilmiştir. Sismik kesitler incelenirken, Kretase yaşlı birimlerin içerisinde birçok küçük ölçekli normal faylar ayırt edilmiştir. Bu faylar, sismik kesitlerde Kretase yaşlı birimlerin üzerini örten Eosen yaşlı birimlerin içerisinde ayırt edilen dilimli yapıdaki bindirme faylarından (imbricate thrust faults) farklı özellikler göstermektedir.

Sismik kesitlerden yararlanılarak oluşturulan jeolojik kesitlerin kıyıya yakın olan kesimlerinde, Eosen birimleri içerisinde bulunan (sismik kesitlerde Pliyosen sınırından itibaren gözlemlenemeyen) ve dilimli yapı özelliği gösteren bindirme fayları işaretlenmiştir. Bu fayların fay düzlem açıları kuzeye doğru azalmaktadır ki bu özellikleri de fayların tabandan bir sıyrılma fayına (detachment fault, décollement) bağlandıklarını göstermektedir. Sismik kesitlerde sıyrılma fayının bulunduğu düşünülen bölümler, tekrarlı yansımalar tarafından örtüldüğünden dolayı sıyrılma fayının ayırt edilmesi pek mümkün olamamıştır. Dilimli yapı gösteren bindirme faylarının tabanındaki sıyrılma fayının konumu, oluşturulan kara-kıyı ötesi jeolojik kesitlerde, tahmini olarak (kesikli çizgilerle) işaretlenmiştir. Sıyrılma fayı, oluşturulan kara-kıyı ötesi kesitlerinden de görülebileceği gibi, Kretase yaşlı birimlerin içerisinden geçerek Eosen yaşlı birimler içerisinde bulunan bindirme faylarının tabanına doğru uzanmaktadır. 03 Eylül 1968 Bartın Depremi'nin odağına en yakın hat olan 11 numaralı hatta ait kara-kıyı ötesi jeolojik kesit (Şekil 16) incelendiğinde, depremin odak derinliğinin sıyrılma fayının bulunduğu düşünülen derinlikte yer aldığı görülmektedir. Bu sebeple, Bartın Depremi'nin oluşumunun sıyrılma fayındaki bir harekete bağlı olarak gerçekleşmiş olduğu düşünülmektedir.

Sismik kesitlerden yararlanılarak oluşturulan jeolojik kesitler ve önceki çalışmalardan elde edilen jeolojik kesitlerin kullanılmasıyla yorumsal olarak oluşturulan kara-kıyı ötesi jeolojik kesitlerinde (Şekil 12 ve Şekil 16), kıyıya yakın kesimlerde yüzeylemiş olan jeolojik birimlerin çoğunlukla Kretase'den daha yaşlı birimler olduğu dikkat çekmektedir. Bu kesitler, karadaki daha yaşlı birimlerin genç birimler üzerine bindirmesi ve Batı Karadeniz Havzası'nın güney ve güneybatı bölümleri boyunca oluşmuş olan sıkışma, bindirme ve yükselme olaylarının sonuçları hakkında faydalı bilgiler sunmaktadır. Bu çalışmada elde edilen sonuçlar, Karadeniz'in K-G yönlü olarak sıkışmakta olduğu düşüncesini destekler niteliktedir.

## SUMMARY

Geophysical data have become increasingly important for rapidly developed tectonic research since 1960s. Seismic reflection method is the most effectively used geophysical method for investigation of the deep structure and the geological features beneath the sea floor. From the middle of 1990's to present, in spite of well-rounded but non-utilizable data of commercial two dimensional (2D) and three dimensional (3D) seismic studies of petroleum companies, the number of scientific studies and publications for academic purposes, revealing the geological structure beneath Turkish sector of the Western Black Sea Basin using marine seismic reflection data, has been increasing. Especially, there is a considerable amount of scientific articles focused on various subjects (such as mass wasting, sedimentation, sea-level fluctuations, BSR, Messinian events etc.) presenting detailed seismic sections and providing a good definition of the structural properties beneath the shelf and the continental slope of the southern part of the basin along Turkish margin.

In tectonics, researchers search for ways to predict the time and location of damaging tectonic events such as earthquakes and tectonically induced events. To be socially useful, the predictions must be precise enough to alleviate the loss of life and property. This quest for prediction techniques involves both the investigation of neotectonics and active tectonics (Moores and Twiss, 1995). The earthquake occurred on October 15, 2016 ($M_l$=5.0) (KRDAE, 2016) re-attracted attention to the tectonic activity of Western Black Sea Basin. The focal mechanism solution of this earthquake indicated reverse faulting, similar to the Bartın Earthquake of September 3, 1968 ($M_S$=6.6) (Alptekin et al., 1986; Tan, 1996; Taymaz et al., 1999), which is the strongest instrumentally recorded earthquake along the Turkish margin of Black Sea, and revealed another seismological evidence for the active thrusting in the region.

The folds are commonly accompanied by reverse faulting and many of these reverse faults are low angle and are called thrust faults. Some of the thrust faults and high-angle reverse faults may break the surface, but many others remain hidden within the cores of anticlines and are termed buried reverse faults (Keller and Pinter, 2002). Buried active faults present a significant earthquake hazard-that large damaging earthquakes can occur on faults located entirely beneath or within folded rocks. The tips of such faults may be buried at depths of several kilometers, and when they rupture during earthquakes, uplift and folding occur at the surface (Stein and Yeats, 1989; Keller and Pinter, 2002). In this study, the fault structures that are considered to be formed by the effect of compressional tectonic regime and the structures formed by the activities of these faults beneath the shelf and slope areas between the region offshore Akçakoca-Cide at the southern part of the Western Black Sea Basin, revealed by using marine seismic reflection data.

The marine seismic reflection data used in this study were collected by the collaboration of General Directorate of Mineral Research and Exploration (MTA), Cambridge University and Istanbul Technical University (ITU) in September, 1998. R/V MTA Sismik-1 was used to collect data on 14 seismic lines with a total length of 460 km. All necessary and some optional processes of typical marine seismic data processing sequence were applied to raw data to obtain time-migrated seismic sections. From the time-migrated seismic sections, to give examples for thrust fault structures and folds formed by the activities of these faults became possible. These images were beneficial to clarify the presence of the compressional regime effecting the study area (Examples given for Line 4 and Line 11).

Available onland and offshore geological exploratory well data of Akçakoca-1, Akçakoca-2, Ereğli-1, Filyos-1, Bartin-1, Ulus-1, Amasra-1, Cakraz-1 and Gegendere-1 are obtained from General Directorate of Petroleum Affairs (PIGM) and used to correlate with seismic sections. Seismic markers of the time-migrated seismic sections were dated by referencing to Akcakoca-1 and Akcakoca-2 wells.

Also in this study, the land-offshore geological sections prepared by means of the geological sections given by previous geological studies and the information about the continuation of the geological features from land to offshore in the study area was presented (Examples given for Line 4 and Line 11). These geological sections also give quite beneficial new evidences for the presence of the compressional tectonic regime in the study area.

The important findings of this study were to give another clue for the presence of the compressional tectonic regime effecting the basin by using the outcomes of seismic reflection studies and to reveal the continuation of the thrust related geological features from land to offshore. The results obtained, agrees with the opinion that Black Sea is being compressed in N-S direction.

## KATKI BELİRTME

Jeoloji Mühendisliği Bölümü'nden Doç. Dr. Gürsel SUNAL'a teşekkürlerimizi sunarız.

## DEĞİNİLEN BELGELER